\documentclass[lettersize,journal]{IEEEtran}
\usepackage{amsmath,amsfonts,amssymb,amsthm}
\usepackage{array}
\ifCLASSOPTIONcompsoc
\usepackage[caption=false,font=normalsize,labelfon
t=sf,textfont=sf]{subfig}
\else
\usepackage[caption=false,font=footnotesize]{subfi
g}
\fi
\usepackage{textcomp}
\usepackage{stfloats}
\usepackage{url}
\usepackage{verbatim}
\usepackage{graphicx}
\usepackage{balance}
\usepackage{booktabs}
\usepackage{multirow}
\usepackage[ruled, linesnumbered]{algorithm2e}

\newtheorem{definition}{Definition}

\usepackage{pdfpages}
\usepackage{xcolor}
\usepackage{tabularx}
\usepackage{colortbl}
\usepackage[numbers,sort&compress]{natbib}
\usepackage{hyperref}
\usepackage{pdfpages}

\begin{document}

\title{A Hierarchical Framework with Spatio-Temporal Consistency Learning for Emergence Detection in Complex Adaptive Systems}
\author{Siyuan Chen, and Xin Du, and Jiahai Wang,~\IEEEmembership{Senior Member,~IEEE}
\thanks{
  This work is supported by the National Key R\&D Program of China (2018AAA0101203), the National Natural Science Foundation of China (62406083, 62472461, 62072483), and the Guangdong Basic and Applied Basic Research Foundation (2022A1515011690). \textit{(Corresponding author: Jiahai Wang.)} 
  }
\thanks{
  Siyuan Chen is with the School of Computer Science and Cyber Engineering, Guangzhou University, Guangzhou 510006, China (e-mail: chensiyuan@gzhu.edu.cn).
  
  Xin Du is with the Civil Aviation Electronic Information Engineering College, Guangzhou Civil Aviation College, Guangzhou 510403, China (e-mail: duxin@gcac.edu.cn).
  
  Jiahai Wang is with the School of Computer Science and Engineering, Sun Yat-sen University, Guangzhou 510275, China (e-mail: wangjiah@mail.sysu.edu.cn).
  }
}

\markboth{IEEE Transactions on Neural Networks and Learning Systems}%
{Shell \MakeLowercase{\textit{et al.}}: A Sample Article Using IEEEtran.cls for IEEE Journals}

\maketitle

\begin{abstract}
  Emergence, a global property of complex adaptive systems (CASs) constituted by interactive agents, is prevalent in real-world dynamic systems, e.g., network-level traffic congestions. Detecting its formation and evaporation helps to monitor the state of a system, allowing to issue a warning signal for harmful emergent phenomena. Since there is no centralized controller of CAS, detecting emergence based on each agent's local observation is desirable but challenging. Existing works are unable to capture emergence-related spatial patterns, and fail to model the nonlinear relationships among agents. This paper proposes a hierarchical framework with spatio-temporal consistency learning to solve these two problems by learning the system representation and agent representations, respectively. Spatio-temporal encoders composed of spatial and temporal transformers are designed to capture agents' nonlinear relationships and the system's complex evolution. Agents' and the system's representations are learned to preserve the spatio-temporal consistency by minimizing the spatial and temporal dissimilarities in a self-supervised manner in the latent space. Our method achieves more accurate detection than traditional methods and deep learning methods on three datasets with well-known yet hard-to-detect emergent behaviors. Notably, our hierarchical framework is generic in incorporating other deep learning methods for agent-level and system-level detection.
\end{abstract}
\begin{IEEEkeywords}
Emergence detection, complex adaptive systems, self-supervised learning on dynamic graphs, spatio-temporal modeling.
\end{IEEEkeywords}

\section{Introduction}
\begin{figure}[!t]
    \centering
    \includegraphics[width=0.85\columnwidth]{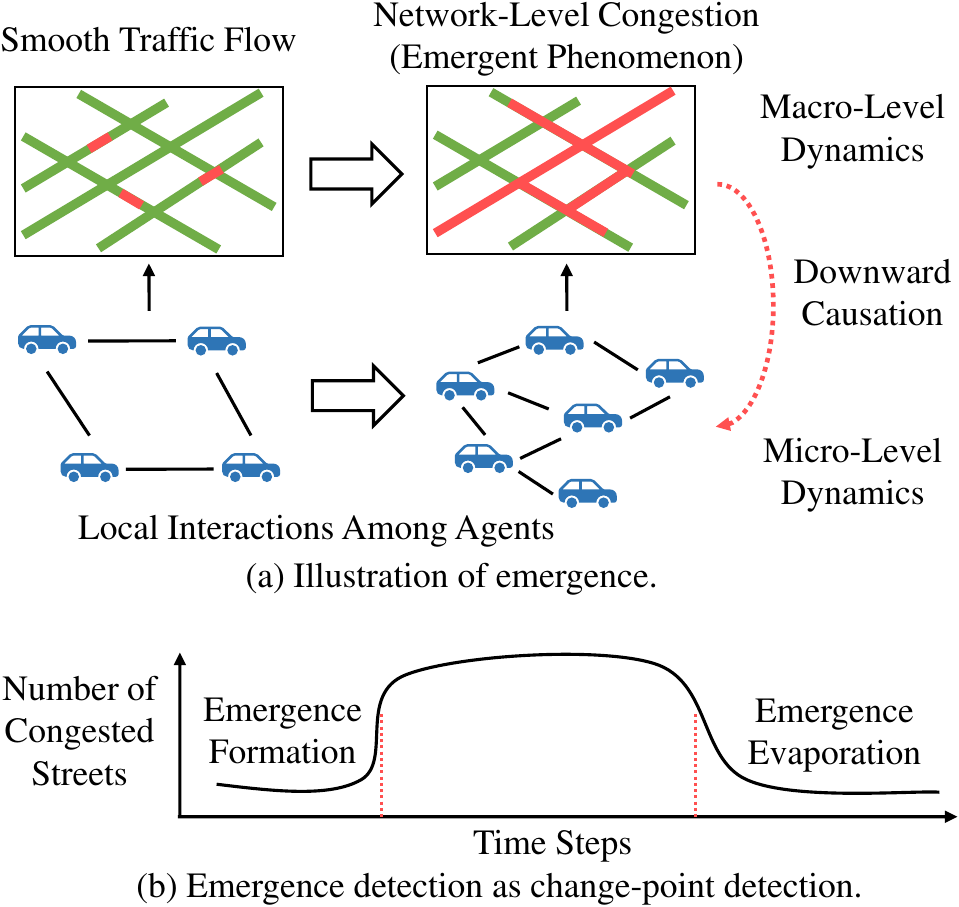}
    \caption{(a) illustrates the emergence through the traffic flow. (b) shows that emergence detection is framed as a change-point detection problem.}
    \label{fig:example}
\end{figure}

Many real-world dynamic systems can be regarded as complex adaptive systems (CASs) composed of autonomous, adaptive, and interacting agents, whose interactions at the micro level can result in emergent phenomena at the macro level, namely, emergence~\cite{artime2022origin,kalantari2020emergence,otoole2016decentralised}. Figure~\ref{fig:example}(a) presents an example. When there is adequate space among cars, the road net enjoys a smooth traffic flow. When the distances between cars significantly narrow down, network-level congestion occurs as an emergent phenomenon. Emergence is irreducible to the properties of agents that constitute CAS, and it is unpredictable by nature~\cite{artime2022origin,fromm2005types}. Nonetheless, it will be beneficial to detect the formation and evaporation of emergence, specifically, weak emergence that is scientifically relevant~\cite{bedau1997weak}. It can help monitor some global properties of the system and issue a warning signal when an undesirable phenomenon arrives. For example, reporting a traffic jam based on the feedback of cars can help with the health management of traffic systems. It will complement existing monitors relying on static devices like sensors or cameras~\cite{zeng2021percolation}.

Emergence detection can be formulated as an online change-point detection (CPD) problem by regarding the time steps when emergence forms or evaporates as change points~\cite{otoole2017decentralised,aminikhanghahi2016survey}. As shown in Figure~\ref{fig:example}(b), the number of congested streets can serve as a global variable to monitor the emergence. However, CASs are distributed by nature, i.e., there is no centralized controller that can access the states of all agents. Therefore, methods requiring all agents' states to compute a global metric for emergence detection become impractical under the distributed setting. Hence, it is desirable to detect emergence using agents' local observation, which is feasible because all agents experience the downward causation~\cite{otoole2017decentralised} when the emergence forms, as shown in Figure~\ref{fig:example}(a). For example, cars slow down in a crowded street. Based on this observation, \citet{otoole2017decentralised} propose DETect, the only feasible emergence detection method for the distributed setting. In DETect, each agent analyzes its relationship with neighbors, communicates with them, and sends feedback when finding a noticeable change in the relationship. DETect concludes the formation or evaporation of emergence when the number of feedback gets significantly larger.

Though making a big step, DETect has two main limitations. First, it simply counts agents' feedback to monitor the emergence, which may neglect the spatial patterns that are highly correlated to emergence. Second, it adopts linear regression to model an agent's relationship with its neighbors, which may fail to capture nonlinear relationships.

This paper tries to overcome the above limitations from a graph perspective. CAS can be regarded as a dynamic graph~\cite{gignoux2017emergence}, and thus emergence detection can be cast to online CPD in dynamic graphs under the distributed setting. Based on this formulation, it suffices to learn a system-level representation aware of emergence-related patterns, and agent-level representations encoding the nonlinear relationships. When emergence forms or evaporates, the system representations between adjacent time steps are expected to be inconsistent, which can serve as a detecting signal for emergence. Specifically, a \textbf{H}ierarchical framework with \textbf{S}patio-\textbf{T}emporal \textbf{C}onsistency \textbf{L}earning is proposed for emergence detection (\textbf{HSTCL}). HSTCL is of a three-layer structure, \textit{``agents-region monitors-global monitor''}, which allows to capture emergence-related spatial patterns by aggregating agent-level detecting results from bottom-up. HSTCL can be conceptually implemented by the efficient end-edge-cloud collaborative framework. Spatio-temporal encoders (STEs) composed of spatial and temporal transformers are designed to model the complex variation of agents' nonlinear relationships and the system states in highly dynamic scenarios. Representations of agents and the system are learned to jointly preserve the spatial and temporal consistency by respectively minimizing the spatial and temporal dissimilarities in the latent space. Compared with DETect, HSTCL can capture non-linear spatio-temporal relationships with the aid of STEs, and identify system-wide emergence-related spatial patterns beyond agents' scope. As a framework, HSTCL is more flexible than DETect. It can incorporate other deep learning methods to further boost the performance. Our contributions are summarized as follows.
\begin{itemize}
    \item The hierarchical framework HSTCL can capture emergence-related spatial patterns by aggregating agent-level detecting results from bottom-up.
    \item STEs composed of spatial and temporal transformers are designed to model the nonlinear relationships among agents and the evolution of the system. Featured by parallel execution and incremental update of representations, these encoders are especially suitable for online detection.
    \item The agent representations and the system representation are learned to preserve the intrinsic spatio-temporal consistency in a self-supervised manner. The training strategy avoids data augmentations that may break spatio-temporal consistency and negative samples that increase the computational overhead.
    \item Extensive experiments on three datasets with well-known yet hard-to-detect emergent phenomena validate the superiority of HSTCL over DETect and deep learning methods. Notably, other deep learning methods can be incorporated in our framework for emergence detection.
\end{itemize}

\section{Related Work}

\subsection{Emergence Detection}
Detecting the emergence of CAS, generally requires at least one global monitor~\cite{kalantari2020emergence}. Depending on how the monitor acquires the information of agents for detection, existing methods fall into three design choices of architectures: \textbf{I)} a single monitor with direct access to agents' states; \textbf{II)} a monitor collecting agents' information indirectly, e.g., from static sensors ; \textbf{III)} a monitor collecting feedback from mobile agents. Our method belongs to class III.

Class I methods define global variables to monitor the system state, e.g., information entropy~\cite{mnif2011quantitative,prochazka2015monitoring,liu2018mechanism,luo2022dynamic}, statistical divergence~\cite{fisch2010quantitative}, and Granger-emergence~\cite{seth2008measuring}. These methods require centralized monitoring, and are thus inapplicable under the distributed setting. Class II methods allow distributed monitoring. However, they requires prior knowledge of emergence to decide what to detect at each sensor~\cite{grossman2008discovering,niazi2011sensing}, falling short in detecting unknown emergent phenomena. DETect~\cite{otoole2017decentralised}, the only existing method from class III, overcomes the limitations of the first two classes. Each agent serves as a local detector, and agents' feedback is aggregated to monitor the emergence. Our method inherits the advantages of DETect, and further introduces region monitors between agents and the global monitor, allowing to analyze spatial patterns ignored by DETect. STEs are tailored to model nonlinear relationships among agents, which is difficult for the linear models adopted by DETect.

Similar to region monitoring, \citet{santos2017automatic} detect emergence by utilizing subsystem-level information. Their work requires collecting and labeling data of pre-defined subsystems, which is not applicable to emergence detection rooted in agents' local observations. More backgrounds of CAS and emergence, along with a detailed description of DETect are shown in \textit{Appendix A}.

\subsection{Related Graph Mining Tasks}
Emergence detection can be viewed as CPD in dynamic graphs~\cite{huang2024laplacian,sulem2024graph}. It is also closely related to graph-level anomaly detection (AD)~\cite{ma2021comprehensive,pazho2024survey} and multivariate time series AD~\cite{deng2021graph,zheng2023correlation}, since emergence is a novel global property. Structural changes from the perspectives of edges~\cite{wang2017fast}, single-view~\cite{huang2020laplacian} and multi-view~\cite{huang2024laplacian} graph Laplacian have been sustainably explored for CPD. Finer-grained detection is also studied on overlapping communities w.r.t. different stages of evolution~\cite{cheng2020overlapping}. 
Nonetheless, these methods cannot jointly model the changes in node features and structures. Graph neural networks (GNNs)~\cite{li2024guest} can overcome this limitation. sGNN~\cite{sulem2024graph} adopts siamese GNNs to compare the similarity between two adjacent graph snapshots, but it ignores the temporal dependence of graph snapshots. An offline detection method~\cite{zhang2024vggm} uses the Gaussian mixture model to cluster the graph snapshots and identifies a change point when adjacent graph snapshots belong to different clusters. However, it is unsuitable for online detection because it needs to acquire the information of graphs over all time steps. 

For graph-level AD, the structural changes of dynamic networks are studied from the levels of nodes, communities, and the full-graph~\cite{jiao2021generative,jiao2023hbdsbm}. Related multi-scale dependence is also explored via graph framelets~\cite{li2024permutation} and graph contrastive learning~\cite{zheng2024selfsupervised} in general graph learning methods. For time series AD, the anomaly-related multi-scale spatio-temporal patterns are modeled by dilated temporal convolution and multi-hop GNNs~\cite{zheng2023correlation}, and the cross-time spatial dependence is modeled by the fuzzy embedding~\cite{zhu2024fuzzy}. The anomaly is measured by prediction error~\cite{deng2021graph,zheng2023correlation}, one-class classification loss~\cite{zhao2021using}, contrastive loss~\cite{qiu2022raising}, etc. However, these methods are originally designed for centralized detection. \citet{protogerou2021graph} propose a distributed graph anomaly detection method, where each node shares the latent vector with its neighbors. This will raise privacy concerns and increase the communication cost. Thus, methods from graph-level CPD and AD are inapplicable for emergence detection, but they can adapt to our framework regarding the distributed settings~\cite{otoole2017decentralised}, where agents can only sense their neighbors' states and share the detecting results.

\subsection{Self-Supervised Learning for Spatio-Temporal Data}
Rich deep learning methods that capture multi-scale spatio-temporal correlations have been developed for spatio-temporal data like videos~\cite{qin2023coarse} and dynamic graphs~\cite{hu2022spatio}, with applications to spatio-temporal forecasting~\cite{hu2022spatio}, task scheduling~\cite{yuan2019spatiotemporal}, decision-making~\cite{wang2020stmarl}, etc. However, the scarcity of labels makes it difficult to effectively train complex models. Self-supervised learning is explored to leverage rich information underlying the unlabeled data, with success in time series~\cite{zhang2024selfsupervised}, videos~\cite{schiappa2023self} and static graphs~\cite{wu2023self,zheng2024selfsupervised}. However, the efforts in dynamic graphs are limited. Contrastive learning is a typical paradigm, but it is non-trivial to construct different views of a node or a graph that preserve spatio-temporal semantics~\cite{liu2022contrastive,zhang2023automated}. Besides, the large number of negative samples substantially increases the computational overhead. To avoid these issues, this paper develops non-contrastive spatio-temporal learning strategies.

\section{Proposed Method}
\begin{figure}[!t]
    \centering
    \includegraphics[width=0.6\columnwidth]{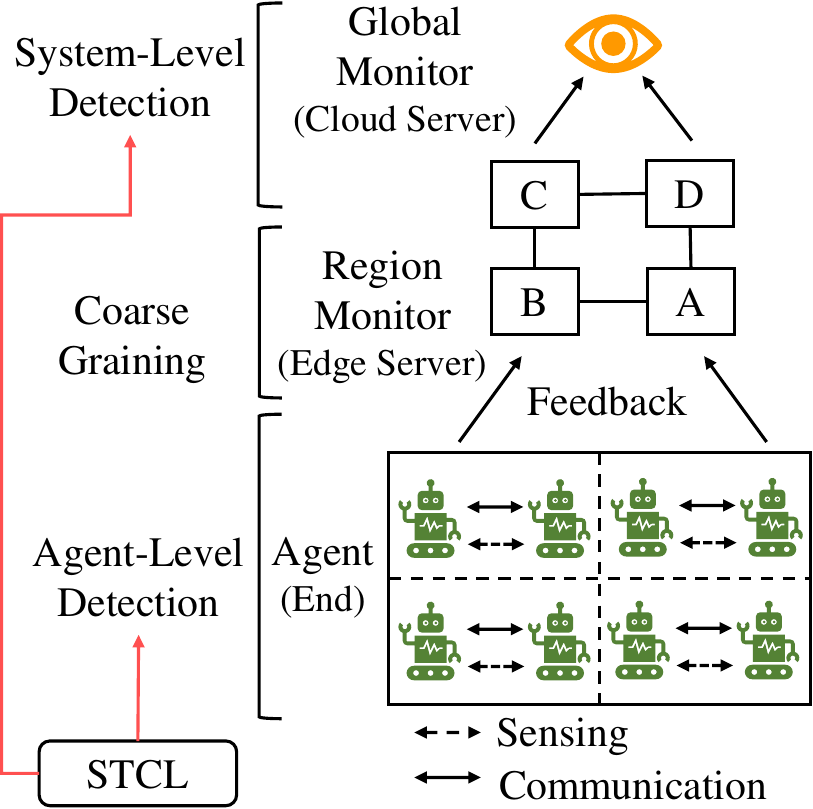}
    \caption{Overview of HSTCL. HSTCL contains three hierarchies, agents, region monitors, and a global monitor, which can be conceptually implemented by the end-edge-cloud collaborative framework. Agents sense the states of neighbors, measure the change in relationships, and communicate with them to make agent-level detection. The detecting results are coarse-grained to region states, whose spatial patterns are used for system-level detection. Representations of agents and the system are learned via spatio-temporal consistency learning (STCL) technique to support agent-level and system-level detection, respectively.}
    \label{fig:overview}
\end{figure}
\subsection{Problem Formulation}\label{sec:method:problem_formulation}
Regarding the time steps when the emergence forms or evaporates as change points, emergence detection in CAS can be formulated as CPD in dynamic graphs as follows.
\begin{definition}[Dyanmic graph]
    A dynamic graph is composed of a graph series $\{\mathcal{G}^t\}^T_{t=1}$, where $\mathcal{G}^t=(\mathcal{V}, \mathcal{E}^t, \mathbf{X}^t)$ is a snapshot at time $t$, with $\mathcal{V}$ as the set of nodes over all time steps, $\mathcal{E}^t$ as the set of edges at time $t$, and $\mathbf{X}^t$ as the node features at time $t$.
\end{definition}

\begin{definition}[CPD in dynamic graphs]
    The task of CPD in dynamic graphs aims to find a set of change points $\mathcal{T}^*=\{t^*_k\}^{K}_{k=1}$ from a graph series $\{\mathcal{G}^t\}^T_{t=1}$. The change points split $[1, T]$ into contiguous segments such that $[1, T]=\bigcup^{K^*-1}_{k=1}[t^*_k, t^*_{k+1}]$, with $t^*_1=1$ and $t^*_{K}=T$.
\end{definition}

The node feature $\mathbf{x}^t\in\mathbb{R}^4$ is an agent's 2D position and velocity. The topology of graphs can change over time, including the addition and removal of nodes and edges. For convenience, this paper groups all appearing nodes and assumes the node set is time-invariant, i.e.,  $\mathcal{V}=\cup^T_{t=1}\mathcal{V}^t$~\cite{huang2024laplacian}. The edge set $\mathcal{E}^t$ is essentially dynamic because the edge is defined by thresholding the distance between two agents, as will be described in Definition~\ref{def:distributed_setting}. Specifically, the curve of the number of edges over time is reported in \textit{Appendix D}.

Under the distributed setting, the global monitor does not have direct access to agents' states. Each agent as a local detector has limited vision and shares limited messages.
\begin{definition}[Distributed Setting for Emergence Detection]
A qualified emergence detection method under the distributed setting should satisfy the following three conditions:
    \begin{itemize}
        \item \textbf{Condition 1.} An agent only senses the states of other agents within a certain radius. Formally, the neighborhood of agent $j$ at time $t$ is defined as $\mathcal{N}^t_j=\{i:d^t_{ij}\leq \delta\}$, where $d^t_{ij}$ is the Euclidean distance.
        \item \textbf{Condition 2.} An agent $j$ only communicates with its neighbors in $\mathcal{N}^t_j$.
        \item \textbf{Condition 3.} The only message that an agent $j$ shares with its neighbors or uploads to some monitor is its detecting score for emergence, i.e., a scalar $s^t_j\in [0, 1]$.
    \end{itemize}
    \label{def:distributed_setting}
\end{definition}

\noindent Inspired by \citet{stephen2015anomaly}, this paper uses a dissimilarity function to calculate the detecting scores, and defines the criterion for CPD as follows.
\begin{definition}[Criterion for CPD]
    Given a graph series and a dissimilarity function $d(\mathcal{G},\mathcal{G}')\in\mathbb{R}_{\geq 0}$, a change point $t$ is detected when $d(\mathcal{G}^t,\mathcal{G}^{t-1})>c$ and $d(\mathcal{G}^t,\mathcal{G}^{t+1})\leq c$ for some threshold $c$.
    \label{def:cpd_criterion}
\end{definition}

\subsection{Motivation and Overview of HSTCL}
The distributed setting stated in Definition~\ref{def:distributed_setting} poses severe challenges to existing spatio-temporal modeling techniques and emergence detection methods:
\begin{itemize}
    \item [(1)] \textit{Conditions 1 and 3} state that the hidden vectors of agents are not shared. Thus, the common practice of stacking multiple GNN layers~\cite{wu2020comprehensive} to capture long-range dependence over multi-hop neighbors is inapplicable.
    \item [(2)] \textit{Conditions 2 and 3} state that it is hard to reach consensus among agents via local communication within a limited time, because the communication graph changes constantly and it is not necessarily connected~\cite{ren2008distributed}. See \textit{Appendix A} for further demonstration.
    \item [(3)] \textit{Condition 3} states that the global monitor is unable to make global detection by utilizing agents' states in an end-to-end manner.
\end{itemize}

These challenges motivate the key design choice of HSTCL, that is, modeling the spatio-temporal dependency at different levels and aggregating the information hierarchically. An overview of HSTCL is shown in Figure~\ref{fig:overview}. It contains three hierarchies from bottom-up, \textit{agents, region monitors, and a global monitor}. It can be conceptually implemented by the end-edge-cloud collaborative framework~\cite{he2022pyramid}. The area where all agents move is split into several connected regions. In each region, every agent senses the states of its neighbors and detects if its relationship with neighbors changes significantly. Each agent communicates with its neighbors to enhance the agent-level detecting results. The detecting results of agents within the same region are aggregated by the corresponding region monitor. The regional results are analyzed by the global monitor to make a system-level detection that is aware of emergence-related patterns.

Agent-level detection compresses an agent's local observations into a single detection score, while system-level detection unifies these scores to gain a global view. They are supported by agent-level and system-level representation learning, respectively. STEs are designed to capture nonlinear agent-to-agent and region-to-region relationships. Spatio-temporal consistency learning (STCL) strategy guides STEs to learn agents' and the system's representations that preserve both spatial and temporal consistency. Both agent-level and system-level STCL preserve the temporal consistency of representations within a time window. The former preserves the spatial consistency between each agent's and its neighbors' representations, and the latter preserves the spatial consistency between the system's and the regions' representations. The inconsistency in system representation serves as a detection signal for emergence. Formally, HSTCL can be described as a three-step process, corresponding to its three-level structure,
\begin{equation}
    \begin{aligned}
        \mathbf{s}^{1:T} &= \text{AgentDetect}\left(\mathbf{X}^{1:T}\right)\in\mathbb{R}^{T\times |\mathcal{V}|},\\
        \mathbf{y}^{1:T} &= \text{CoarseGrain}\left(\mathbf{s}^{1:T}\right)\in\mathbb{R}^{T\times M},\\
        s^{1:T}_\mathcal{G} &= \text{SystemDetect}\left(\mathbf{y}^{1:T}\right)\in\mathbb{R}^{T}.
    \end{aligned}
    \label{eq:hstl_framework}
\end{equation}
$\mathbf{s}^{1:T}$ are agent-level detecting scores, which are coarse-grained to $M$ regions' states $\mathbf{y}^{1:T}$. $s^{1:T}_\mathcal{G}$ are system-level detecting scores based on region states. The following sections will describe the process of HSTCL in detail.

\subsection{Agent-Level Detection}
\subsubsection{Spatio-Temporal Encoder} To make online detection at time $\tau$, each agent records its state and its neighbors' states in the last $w$ time steps. Denoting $\tau(w)=\tau-w+1$ as the initial step of the time window, these states are transformed to agent representations by the agent-level STE,
\begin{equation}
    \mathbf{h}^{\tau(w):\tau}_j =\text{STE}_A\left(\mathbf{x}^{\tau(w):\tau}_j,\left\{\mathbf{x}^{t}_i:i\in\mathcal{N}^t_j\right\}^\tau_{t=\tau(w)}\right)\in\mathbb{R}^{w\times D},
    \label{eq:agent_spatial_transformer}
\end{equation}

Due to \textit{Conditions 1 and 3} of the distributed setting, each agent cannot acquire their neighbors' latent representations. Thus, the popular design choice of integrated dynamic GNNs that model spatio-temporal entangled relations~\cite{skarding2021foundations,wang2022deep} is inapplicable. Instead, this paper adopts a stacked architecture composed of a spatial transformer and a temporal transformer~\cite{vaswani2017attention}, disentangling spatial and temporal dependency. 

The spatial transformer models the relationship between an agent and its neighbors at each time step. The state of an agent $\mathbf{x}^t_j$ is first embedded as a hidden vector through a single-layer perceptron, i.e., $\mathbf{e}^t_j=\text{Emb}(\mathbf{x}^t_j)$. Then, a scaled dot-product attention mechanism~\cite{vaswani2017attention} with a skip connection is applied to calculate the spatial representation
\begin{equation}
    \mathbf{z}^t_j = \mathbf{e}^t_j + \sum_{i\in\mathcal{N}^t_j} \alpha^t_{ij} f_V\left(\mathbf{e}^t_i - \mathbf{e}^t_j\right).
\end{equation}
$\mathbf{e}^t_i - \mathbf{e}^t_j$ measures the spatial difference between agent $j$ and its neighbor $i$ in the latent space. It may capture nonlinear relations that are not fully reflected in quantities of the raw space, e.g., relative positions and velocities. $f_V$ is a value function implemented by a linear mapping. $\alpha^t_{ij}=\text{softmax}(\{a^t_{ij}:i\in\mathcal{N}^t_j\})$ is the normalized attention score, with $a^t_{ij}$ defined as
\begin{equation}
    a^t_{ij} = \frac{1}{\sqrt{D}}f_Q\left(\mathbf{e}^t_j\right)^\top f_K\left(\mathbf{e}^t_i\right),
\end{equation}
where $f_Q$ and $f_V$ are linear layers accounting for the query function and the key function, respectively. The spatial transformer does not require temporal embeddings of neighbors within a time window, which is desirable because the neighbors frequently change.

The temporal transformer is instantiated as a standard transformer~\cite{vaswani2017attention}, because it is powerful for sequential modeling,  and it allows parallel execution, which is favorable for online detection. At its core is a temporal attention mechanism that maps spatial representations to temporal representations,
\begin{equation}
    \mathbf{h}^{\tau(w):\tau}_j =\text{softmax}\left(\frac{1}{\sqrt{D}}\mathbf{q}^{\tau(w):\tau}_j\left(\mathbf{k}^{\tau(w):\tau}_j\right)^\top\right)\mathbf{v}^{\tau(w):\tau}_j,
    \label{eq:temporal_transformer}
\end{equation}
where $\mathbf{q}^{\tau(w):\tau}_j, \mathbf{k}^{\tau(w):\tau}_j$ and $\mathbf{v}^{\tau(w):\tau}_j$ are query, key and value vectors transformed from $\mathbf{z}^{\tau(w):\tau}$, respectively.

Disentangling the spatial and temporal information also makes the spato-temporal encoder friendly to streaming data because the agent representations can be updated incrementally. As the time step $\tau$ increases by 1, only the current spatial representation $\mathbf{z}^{\tau+1}_j$ needs to be computed, while $\mathbf{z}^{\tau(w)+1:\tau}_j$ can be reused. For the temporal transformer, intermediate results like the unnormalized attention scores and the query vectors that only involve $\mathbf{z}^{\tau(w)+1:\tau}_j$ can be stored. The normalized attention scores and the temporal representations can be computed by further incorporating $\mathbf{z}^{\tau+1}_j$. Details can be found in \textit{Appendix C}.
\subsubsection{Spatio-Temporal Consistency Learning}
\begin{figure}[!t]
    \centering
    \includegraphics[width=0.98\columnwidth]{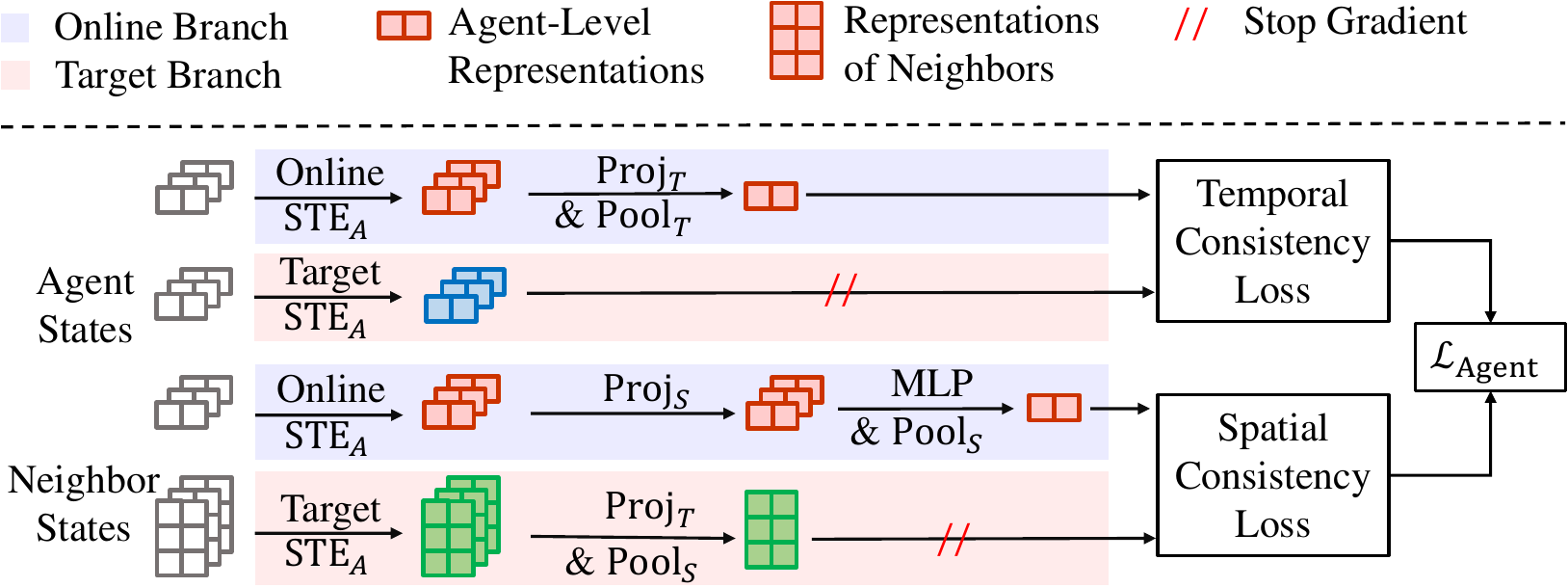}
    \caption{Procedure of agent-level STCL.}
    \label{fig:agent_stcl}
\end{figure}
Since the labels of emergence are rarely known a priori, this paper proposes to train STE in a self-supervised manner by preserving the spatio-temporal consistency of the agent representations. STCL is inspired by an influential non-contrastive method called bootstrapping your own latent (BYOL)~\cite{grill2020bootstrap}, which avoids explicit negative samples by aligning different views of the same sample encoded by asymmetric neural networks. BYOL is briefly introduced in \textit{Appendix A}.

Unlike BYOL that uses a single objective, STCL disentangles the learning objectives of temporal consistency and spatial consistency since they characterize different aspects of the dynamic system. For each aspect, an online network and a target network with asymmetric network structures are designed to process different views of the same agent. These views are constructed by leveraging the intrinsic spatial and temporal relations within the data other than data augmentation that may damage the spatio-temporal semantics~\cite{liu2022contrastive}. Given a view of some agent, the online network is trained to align the output of the target network for another view. The procedure of agent-level STCL is depicted in Figure~\ref{fig:agent_stcl}.

In the following subsections, a symbol with a tilde stands for an element from the target branch, e.g., a vector $\widetilde{\mathbf{h}}$ and a function $\widetilde{f}$. Symbols without a tilde come from the online branch. A vector with a superscript $t$ is called a transient representation at time $t$, e.g., $\mathbf{h}^t$. A vector with a superscript $(\tau)$ stands for a short-term representation within a time window $[\tau(w),\tau]$, e.g., $\mathbf{h}^{(\tau)}$.

\paragraph{Temporal Consistency Loss}
When emphasizing the temporal consistency, the agent representation $\mathbf{h}^t_j$ is mapped to a temporal space via the temporal projection $\text{Proj}_T$. The resulting vectors $\mathbf{v}^t_j$ are reduced to a short-term representation via a temporal pooling function $\text{Pool}_T$, mean pooling here:
\begin{equation}
        \mathbf{v}^{t}_j = \text{Proj}_T\left(\mathbf{h}^t_j\right),\quad\mathbf{v}_j^{(\tau)} = \text{Pool}_T\left(\mathbf{v}^{\tau(w):\tau}_j\right).
\end{equation}
To ensure that the short-term representation is consistent with the transient representations, and thus capturing the tendency within the time window, this paper minimizes the dissimilarity between them. The dissimilarity is defined as the complement of the cosine similarity,
\begin{equation}
    d\left(\mathbf{v}_j^{(\tau)},\widetilde{\mathbf{h}}^t_j\right) = \frac{1}{2}\left(1 - \text{cos}\left(\mathbf{v}_j^{(\tau)},\widetilde{\mathbf{h}}^t_j\right)\right).
\end{equation}
Then, the temporal consistency loss is defined as the average temporal dissimilarity of all agents within the time window,
\begin{equation}
    \mathcal{L}_T = \frac{1}{w|\mathcal{V}|}\sum_{j\in\mathcal{V}}\sum^\tau_{t=\tau(w)}d\left(\mathbf{v}_j^{(\tau)},\widetilde{\mathbf{h}}^t_j\right).
\end{equation}

\paragraph{Spatial Consistency Loss}
When emphasizing the spatial consistency, $\mathbf{h}^t_j$ is mapped to a spatial space via the spatial projection $\text{Proj}_S$. To avoid disturbing the optimization of the temporal counterpart, this paper further transforms the resulting vectors with a multi-layer perceptron (MLP) to construct an asymmetric branch, i.e.,
\begin{equation}
            \mathbf{n}^{t}_j = \text{Proj}_S\left(\mathbf{h}^t_j\right),\quad
            \mathbf{m}_j^{(\tau)} = \text{Pool}_T\left(\text{MLP}\left(\mathbf{n}^{\tau(w):\tau}_j\right)\right).
\end{equation}

By minimizing the dissimilarity between the short-term representation of each agent and its neighbors, the model learns to preserve spatial consistency, i.e.,
\begin{equation}
\mathcal{L}_S = \frac{1}{\kappa|\mathcal{V}|}\sum_{j\in\mathcal{V}}\sum_{i\in\mathcal{N}_j}d\left(\mathbf{m}_j^{(\tau)},\widetilde{\mathbf{n}}_i^{(\tau)}\right),
\end{equation}
where $\mathcal{N}_j$ contains $\kappa$ random neighbors from the temporal neighborhood $\cup^\tau_{t=\tau(w)}\mathcal{N}^t_j$. The sampling probability of a neighbor is proportional to its frequency.

As $\text{STE}_A$ is responsible for representation, while $\text{Proj}_T$ and $\text{Proj}_S$ are responsible for projections, they are simply implemented as MLPs. Mean pooling is adopted for $\text{Pool}_T$ and $\text{Pool}_S$ for simplicity, and more advanced spatial pooling~\cite{wu2020comprehensive,grattarola2024pooling} and temporal pooling~\cite{lee2021learnable} methods are left for future work.

\paragraph{Optimization} 
Combining the temporal consistency loss and the spatial consistency loss, the overall loss for agent-level learning is
\begin{equation}
    \mathcal{L}_{\text{Agent}} = \mathcal{L}_T + \mathcal{L}_S.
\end{equation}
Directly minimizing the above loss will lead to collapsed representations~\cite{grill2020bootstrap}. To avoid this, the parameters $\Theta_A$ of the online branch are optimized by a gradient-based algorithm, e.g., Adam~\cite{adam}, while parameters $\widetilde{\Theta}_A$ of the target branch are updated by exponential moving average~\cite{grill2020bootstrap},
\begin{equation}
        \Theta_A \leftarrow \text{Opt}\left(\mathcal{L}_{\text{Agent}},\Theta_A\right),\quad
        \widetilde{\Theta}_A \leftarrow \eta \widetilde{\Theta}_A + (1 - \eta)\Theta_A,
    \label{eq:update_parameters}
\end{equation}
where $\eta\in [0, 1]$ is a decay rate.
The final $\Theta_A$ for emergence detection is obtained when the iterative process converges.

\subsubsection{Communication}
Although each agent can make detection independently, sharing the detecting scores will make the detection more robust. In DETect~\cite{otoole2017decentralised}, an agent only communicates with a randomly selected neighbor. Taking a step further, our method allows each agent to update its own score $s^t_j$ by combining the scores of all neighbors and the dissimilarity between representations of adjacent steps,
\begin{equation}
    s^{\tau+1}_j = \alpha \cdot d\left(\mathbf{h}^{(\tau)}_j,\mathbf{h}^{(\tau-1)}_j\right) + \frac{(1-\alpha)}{|\mathcal{N}_j^\tau|+1}\sum_{i\in\mathcal{N}^\tau_j\cup\{j\}}s^\tau_i.
    \label{eq:communication}
\end{equation}
where $\mathbf{h}^{(\tau)}_j=p_T\left(\mathbf{h}^{\tau(w):\tau}\right)$. $\alpha\in[0, 1]$ is a mixing coefficient. When $\alpha=1$, the messages from neighbors are ignored, and when $\alpha=0$, the agent is overwhelmed by its neighbors' detecting scores. The communication cost can be controlled by setting a budget for the number of neighbors.
\begin{figure}[!t]
    \centering
    \includegraphics[width=0.98\columnwidth]{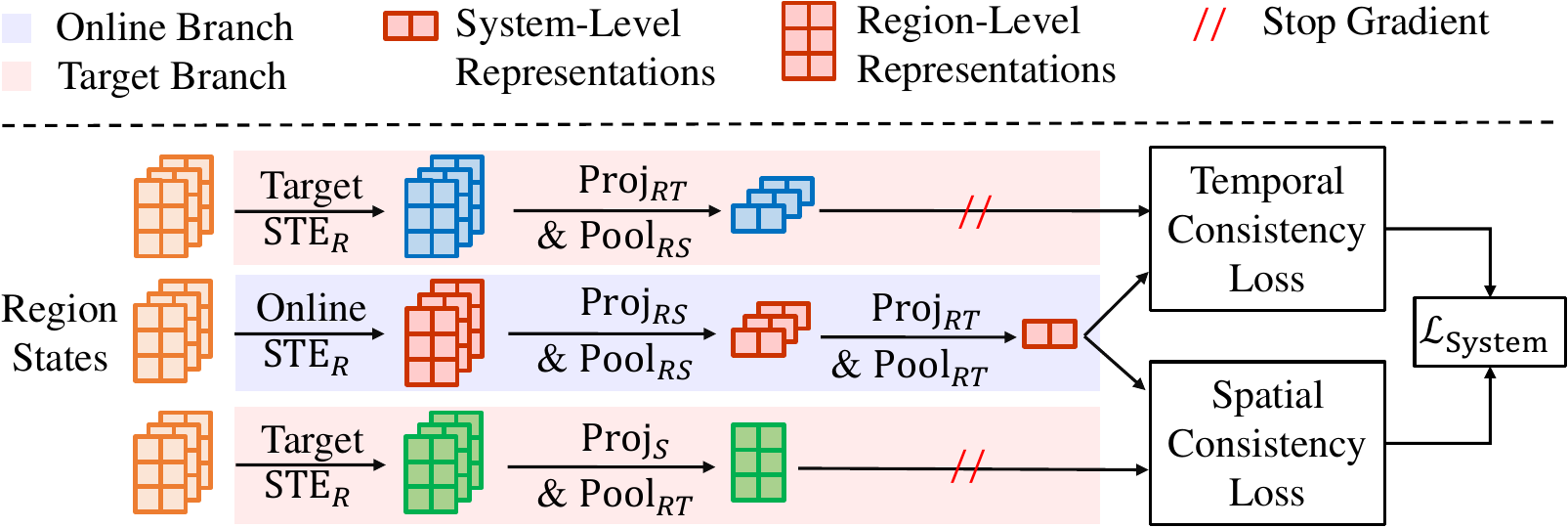}
    \caption{Procedure of system-level STCL.}
    \label{fig:system_stcl}
\end{figure}
\subsection{Coarse Graining and System-Level Detection}
\subsubsection{Coarse Graining}\label{sec:method:coarse_graining}
The area where all agents move is split into several adjacent regions $\{\mathcal{R}_m\}^M_{m=1}$ in a grid shape. The detecting scores of agents within a region are aggregated as the region's state,
\begin{equation}
    y^t_m = \sum_{j\in\mathcal{R}_m}s^t_j.
\end{equation}
These regions form a region graph $\mathcal{RG}^t=(\mathcal{RV},\mathcal{RE},\mathbf{y}^t)$, with $\mathcal{RV}$ as the set of regions, $\mathcal{RE}$ as the set of edges between regions, and $\mathbf{y}^t$ as the region states at time $t$. The formulation can be naturally extended to regions with irregular boundaries and complex graph structures~\cite{sun2020predicting,li2022crowd}. This paper considers grid-shape region graphs for a proof of concept, while more complex scenarios are left for future work.

\subsubsection{Region Representation} As in agent-level detection, a region-level STE with a similar network structure is applied to the region graph for obtaining the representation $\mathbf{r}^t_m$ for each region, i.e.,
\begin{equation}
    \mathbf{r}^{\tau(w):\tau}_m =\text{STE}_R\left(y^{\tau(w):\tau}_m,\left\{y^{t}_n:n\in\mathcal{RN}_m\right\}^\tau_{t=\tau(w)}\right),
    \label{eq:region_spatial_transformer}
\end{equation}
where $\mathcal{RN}_m$ is the set of region $m$'s neighboring regions. Based on region representations, a system representation is learned to gain a global view of the system. Hence, the variation in system representations indicates the formation or evaporation of emergence. Likewise, temporal and spatial consistency losses are designed to guide the learning procedure.

\subsubsection{Spatio-Temporal Consistency Learning}
\paragraph{Temporal Consistency Loss}
The procedure of system-level detection is depicted in Figure~\ref{fig:system_stcl}. A regional spatial projection $\text{Proj}_{RS}$ with a regional spatial pooling function $\text{Pool}_{RS}$ is applied to obtain the transient system representation,
\begin{equation}
        \mathbf{r}^t_\mathcal{G} = \text{Pool}_{RS}\left(\left\{\text{Proj}_{RS}\left(\mathbf{r}^{t}_m\right):m\in\mathcal{RV}\right\}\right).
\end{equation}
Then, a regional temporal projection $\text{Proj}_{RT}$ followed by a regional temporal pooling function $\text{Pool}_{RT}$ is applied to obtain the short-term system representation,
\begin{equation}
    \mathbf{u}^{(\tau)} = \text{Pool}_{RT}\left(\text{Proj}_{RT}\left(\mathbf{r}^{\tau(w):\tau}_\mathcal{G}\right)\right).
\end{equation}

By minimizing the dissimilarity between $\mathbf{u}^{(\tau)}$ and $\widetilde{\mathbf{r}}^t_\mathcal{G}$, the model learns to preserve system-level temporal consistency,
\begin{equation}
    \mathcal{L}_{ST} = \frac{1}{w}\sum^\tau_{t=\tau(w)}d\left(\mathbf{u}^{(\tau)},\widetilde{\mathbf{r}}^t_\mathcal{G}\right).
\end{equation}

\paragraph{Spatial Consistency Loss}
The system-level spatial consistency loss ensures that the system representation $\mathbf{u}^{(\tau)}$ is consistent with the representation of each region. A regional spatial projection $\text{Proj}_{RS}$ together with a regional temporal pooling function $\text{Pool}_{RT}$ is applied to obtain the region representation within a time window,
\begin{equation}
        \widetilde{\mathbf{w}}_m^{(\tau)} = \text{Pool}_{RT}\left(\widetilde{\text{Proj}}_{RS}\left(\widetilde{\mathbf{r}}^{\tau(w):\tau}_m\right)\right).
\end{equation}

By minimizing the dissimilarity between $\mathbf{u}^{(\tau)}$ and $\widetilde{\mathbf{w}}_m^{(\tau)}$, the characteristics of each region are preserved in the system representation,
\begin{equation}
    \mathcal{L}_{SS} = \frac{1}{\kappa}\sum_{m\in\mathcal{D}}d\left(\mathbf{u}^{(\tau)},\widetilde{\mathbf{w}}_m^{(\tau)}\right),
\end{equation}
where $\mathcal{D}$ contains $\kappa$ sampled regions from $\mathcal{RV}$. For simplicity, $\text{Proj}_{ST}$ and $\text{Proj}_{SS}$ are implemented as MLPs, while $\text{Pool}_{RT}$ and $\text{Pool}_{RS}$ are mean pooling, as in agent-level detection. The overall loss for system-level learning is the sum of temporal consistency loss and spatial consistency loss
\begin{equation}
    \mathcal{L}_{\text{System}} = \mathcal{L}_{ST} + \mathcal{L}_{SS}.
\end{equation}

The parameters of region-level online and target networks are updated in the same way as Eq.~\eqref{eq:update_parameters}. Currently, agent-level and system-level STCL are trained separately. The reasons are twofold: (1) the construction of region states requires high-quality agent-level detecting scores; (2) it is hard to define meaningful system-level training signal for agent-level models without the truth change points. A joint optimization of the two hierarchies is left for future work.

The system-level detecting score is defined as the dissimilarity between system representations of adjacent time steps,
\begin{equation}
    s^\tau_\mathcal{G}=d\left(\mathbf{u}^{(\tau)},\mathbf{u}^{(\tau-1)}\right).
\end{equation}

A summary of notations used in this paper is shown in \textit{Appendix B}. The pseudo codes for STCL and emergence detection are shown in \textit{Appendix C}.

\subsection{Time Complexity Analysis of HSTCL}
This paper analyzes the time complexity of HSTCL from its implementation within the end-edge-cloud collaborative framework and in a single machine, respectively corresponding to the potential real-world deployment and the actual implementation in our experiments for proof of concept. Their complexities mainly differ in agent-level detection. 

In the end-edge-cloud collaborative implementation, agents accomplish the computation in parallel~\cite{hua2023edge,gong2023edge}. Recall that $D$ is the dimension of the hidden vector. For agent $j$, the time complexity of spatial encoding at time $t$ is $O(|\mathcal{N}^t_j|D^2)$, and the time complexity of temporal encoding within a time window is $O(wD^2+w^2D)$. Thus, the total complexity of spatio-temporal encoding is $O(wD^2+w^2D+\sum^{\tau}_{t=\tau(w)}|\mathcal{N}^t_j|D^2)$. The time complexities for evaluating the temporal consistency loss and the spatial consistency loss are $O(wD^2)$ and $O(\kappa D^2)$, respectively. The time complexity of communication is $O(|\mathcal{N}^t_j|)$ at each time step. Therefore, at both the training and inference stages, the time complexity is linear w.r.t. the number of neighbors, which can be controlled by setting a budget. Hence, the distributed implementation scales well to large-scale systems. 

In the single-machine implementation, the complexity of agent-level detection is relevant to the number of agents. The time complexity of spatial encoding at time $t$ is $O(|\mathcal{E}^t|D^2)$, and the time complexity of temporal encoding within a time window is $O(|\mathcal{V}|(wD^2+w^2D))$. Thus, the total complexity of spatio-temporal encoding is $O(|\mathcal{V}|(wD^2+w^2D)+\sum^{\tau}_{t=\tau(w)}|\mathcal{E}^t|D^2)$. The time complexities for evaluating the temporal consistency loss and the spatial consistency loss are $O(w|\mathcal{V}|D^2)$ and $O(\kappa|\mathcal{V}|D^2)$, respectively. The time complexity of communication is $O(|\mathcal{E}^t|)$ at each time step. Therefore, at both the training and inference stages, the time complexity is linear w.r.t. the number of agents and the number of edges.

In both implementations, the system-level detection is conducted by the global monitor. The time complexity of system-level spatio-temporal encoding is $O(|\mathcal{RV}|(wD^2+w^2D)+|\mathcal{RE}|wD^2)$. The complexities of evaluating system-level temporal consistency loss and spatial consistency loss are $O(wD^2)$ and $O(\kappa D^2)$, respectively. Thus, the complexity of system-level detection is linear w.r.t. the number of regions and the number of edges, which are generally irrelevant to the number of agents. 

In Section~\ref{sec:exp:running_time}, this paper provides a running time analysis that matches the time complexity analysis to verify the scalability of HSTCL.

\subsection{Characteristics of HSTCL}
HSTCL is characterized by the following features.
\begin{itemize}
    \item [(1)] By hierarchically aggregating agent-level detecting results, HSTCL can capture emergence-related spatial patterns ignored by DETect, where agents' feedback is summed up indiscriminately.
    \item [(2)] Thanks to the spatio-temporal disentangled architecture, STE can capture agents' nonlinear relationships under the distributed setting, where popular designs of spatio-temporal integrated GNNs are infeasible.
    \item [(3)] STCL preserves the spatio-temporal consistency within both agent-level and system-level representations. Compared with BYOL, it avoids potentially harmful data augmentations, and can handle multiple objectives in a disentangled way. It is free of negative samples, significantly reducing the computational cost for spatio-temporal data.
    \item [(4)] HSTCL is flexible in integrating other deep learning methods as long as they satisfy the distributed setting described in Definition~\ref{def:distributed_setting}.  Firstly, existing non-distributed AD and CPD methods can be transformed into distributed detectors by adapting them to agent-level and system-level detection separately. Secondly, STEs can be implemented by other spatio-temporally disentangled GNNs. Thirdly, STCL can be replaced by other self-supervised training schemes like contrastive learning. Lastly, other dissimilarity functions and detection criteria can be adopted for emergence detection.
\end{itemize}

\begin{table}[!tb]
    \renewcommand{\arraystretch}{1.3}
      \caption{Statistics of datasets.}
    \centering
      \begin{tabular}{l|ccc}
    \hline
      Datasets & \multicolumn{1}{c}{Flock} & \multicolumn{1}{c}{Pedestrian} & \multicolumn{1}{c}{Traffic} \\
      \hline
      \# Agents & 150   & 382   & 2,522 \\
      \hline
      Shape of grid & $51\times 51$   & $40\times 40$   & $841\times 841$ \\
      \hline
      \# Simulation steps & 50,000 & 50,000 & 60,000 \\
      \hline
      \# Evaluation steps & 1,000 & 1,000 & 1,200 \\
      \hline
      \# Change points & 10    & 10    & $\approx$ 10 \\
      \hline
      \end{tabular}%
    \label{tab:datasets}%
  \end{table}%

\section{Experiments}
\subsection{Datasets}
For a fair comparison, this paper follows DETect~\cite{otoole2017decentralised} and adopts three simulation environments implemented by NetLogo~\cite{tisue2004netlogo} to generate data. These simulators are equipped with well-known yet hard-to-detect emergent phenomena. In all simulators, agents move in a 2D-bounded world composed of patches. The resulting datasets are briefly described as follows.
\begin{itemize}
    \item Flock~\cite{craig1987flocks}: Each agent is a bird. The emergence is the flocking behavior. The objective measure of emergence is the number of patches that contain no birds.
    \item Pedestrian~\cite{prochazka2015monitoring}: Each agent is a pedestrian walking either from left to right or in the opposite position. The emergence is the counter-flow. The objective measure of emergence is the number of lanes formed by pedestrians.
    \item Traffic~\cite{otoole2017decentralised}: Each agent is a car running on the road net of Manhattan, New York City. The road net contains 6,657 junctions and 77,569 road segments. The cars are routed by a routing engine GraphHopper~\cite{graphhopper} based on real-world car records. The emergence happens when a significant number of streets get congested. Thus, the objective measure of emergence is the number of congested road segments.
\end{itemize}

On the Flock and Pedestrian datasets, real-world data is unavailable. Thus, reasonable behavioral rules are designed for agents. On Traffic dataset, the real-world data is combined with simulation rules to mimic agents' behaviors. Visualizations of emergent behaviors on all datasets and more details of the Traffic dataset are shown in \textit{Appendix D}. A summary of important statistics of the datasets is shown in Table~\ref{tab:datasets}. Each dataset contains 20 times of simulations, with 5 times as the training set, 5 times as the validation set, and the rest as the testing set. Following \citet{otoole2017decentralised}, the objective measure is evaluated every 50 steps. The ground truth change points are labeled by running offline CPD algorithms provided by ruptures~\cite{truong2020selective}. Offline CPD algorithms have access to the whole series, and thus the labeled change points are more reliable and accurate. The results are checked manually. It turns out that change points make up no more than 1\% of all evaluation steps. The severe imbalance between change points and normal points further increases the challenge of emergence detection.

A natural question arises: Why not examine the performance of detectors on real-world datasets? To our best knowledge, there is no benchmark based on purely real-world data.  Currently, collecting such data can be difficult in three aspects: 1) the emergent behavior should be properly understood because labeling emergence formation and evaporation requires prior knowledge; 2) contiguously recording all agents' states for a long period can be challenging to the sensing devices; 3) due to some unavailability issues of real data, it is hard to ensure the diversity. The simulators can overcome these limitations, and they may generate potentially diverse and challenging data that are uneasy to collect in practice, helping to evaluate the detector's performance more comprehensively. We will try to construct qualified real-world datasets in the future.

\subsection{Evaluation Metrics}
Due to the unpredictability of emergence, it can be difficult to detect the exact change points. The formation or evaporation of emergence can happen in a short period rather than a specific time step. Therefore, it is reasonable to accept more than one detection around a true change point in practice. In this paper, the detections within a given tolerance $\theta$ are regarded as one true positive (TP), while the rest detections are regarded as false positive (FP), i.e.,
\begin{align*}
    \text{TP}&=\left|\left\{t^*\in\mathcal{T}^*:\exists t\in\widehat{\mathcal{T}},\text{s.t.}\,|t-t^*|\leq \theta\right\}\right|,\\
    \text{FP}&=\left|\left\{t\in\widehat{\mathcal{T}}:\forall t^*\in\mathcal{T}^*,|t-t^*|> \theta\right\}\right|,
\end{align*}
where $\mathcal{T}^*$ and $\widehat{\mathcal{T}}$ are the set of true change points and the set of detected ones, respectively. This paper sets $\theta=20$ for all datasets. Defining the precision $\text{Prec}=\frac{\text{TP}}{\text{TP}+\text{FP}}$ and the recall rate $\text{Rec}=\frac{\text{TP}}{|\mathcal{T}^*|}$, the F1 score can be computed as
\begin{equation*}
    \text{F1}=\frac{2\times \text{Prec} \times \text{Rec}}{\text{Prec} + \text{Rec}}.
\end{equation*}

The F1 score measures the overall accuracy of CPD. This paper further uses the covering metric~\cite{van2020evaluation,arbelaez2010contour} to measure the overlapping degree between the ground truth segments and the detected segments. Let $\mathcal{A}^*$ be the set of ground truth segments $\mathcal{I}^*$, with a similar definition for $\widehat{\mathcal{A}}$ and $\widehat{\mathcal{I}}$. The covering metric is defined as
\begin{equation*}
    \text{Cover}\left(\mathcal{A}^*,\widehat{\mathcal{A}}\right)=\frac{1}{T}\sum_{\mathcal{I}^*\in\mathcal{A}^*}|\mathcal{I}^*|\cdot \underset{\widehat{\mathcal{I}}\in\widehat{\mathcal{A}}}{\text{max}}\,J\left(\mathcal{I}^*,\widehat{\mathcal{I}}\right),
\end{equation*}
where $J\left(\mathcal{I}^*,\widehat{\mathcal{I}}\right)=\frac{\left|\mathcal{I}^*\cap \widehat{\mathcal{I}}\right|}{\left|\mathcal{I}^*\cup \widehat{\mathcal{I}}\right|}$ is the Jaccard index~\cite{jaccard1912distribution} measuring the overlapping degree between two segments.

In the original paper of DETect~\cite{otoole2017decentralised}, the detecting performance is quantitatively evaluated by checking if the number of detected events is significantly larger during the emergent periods than the non-emergent periods. Nonetheless, the deviation between detected change points and the ground truth is not assessed. Therefore, this paper hopes to fill the gap by introducing the two metrics, and push the current research towards more timely emergence detection.

\subsection{Baselines}
To demonstrate the effectiveness of our method, this paper compares it with DETect and some state-of-the-art deep learning methods in closely related fields, including dynamic network CPD method sGNN~\cite{sulem2024graph}, time series CPD method TS-CPP~\cite{deldari2021time}, time series AD method GDN~\cite{deng2021graph}, and graph-level AD method OCGTL~\cite{qiu2022raising}. Advanced techniques like GNNs and contrastive learning are used in these methods. They are adapted to our framework at both agent-level and system-level detection. They are renamed with a suffix ``+H'', short for Hierarchical framework.

\begin{itemize}
    \item DETect: A decentralized method for online emergence detection. Each agent detects the change in relationships between its neighbors and itself via a linear model. The detecting results are aggregated to make a global decision. 
    \item sGNN+H: A dynamic network CPD method that uses siamese GNNs to learn the graph similarity between two graph snapshots. A top-$k$ pooling module is applied to summarize the node-wise distances in the latent space into a graph similarity.
    \item TS-CPP+H: A time series CPD method based on contrastive learning. Temporal convolutional networks~\cite{bai2018empirical} are used for time series encoding, and the similarity between two contiguous time segments is used as the indicator for CPD.
    \item GDN+H: A GNN-based method for multi-variate time series AD. It uses graph attention to capture the relationships between sensors and defines the maximum deviation score for AD.
    \item OCGTL+H: A graph-level AD method that combines one-class classification and neural transformation learning~\cite{qiu2021neural}, an advanced self-supervised learning technique.
\end{itemize}

The codes of all baselines are publicly available. The code of DETect\footnote{https://github.com/viveknallur/DETectEmergence/} is directly applied to our experiments. The code of sGNN\footnote{https://github.com/dsulem/DyNNet}, TS-CPP\footnote{https://github.com/cruiseresearchgroup/TSCP2}, GDN\footnote{https://github.com/d-ailin/GDN} and OCGTL\footnote{https://github.com/boschresearch/GraphLevel-AnomalyDetection} are adapted to our framework. Implementation details of HSTCL are described in \textit{Appendix D}. The threshold $c$ in Definition~\ref{def:cpd_criterion} is decided by maximizing the F1 score on the validation set.

\subsection{Comparison with Baselines}
\begin{table*}[htbp]
    \renewcommand{\arraystretch}{1.3}
      \caption{Detecting performance of different methods. Both the mean value and the standard deviation are reported. The best results are in \textbf{bold}. The improvements are significant ($p$-value $<0.05$). The relative improvements are computed w.r.t. DETect.}
    \centering
        \begin{tabular}{r|cc|cc|cc}
            \hline
            Datasets & \multicolumn{2}{c|}{Flock} & \multicolumn{2}{c|}{Pedestrian} & \multicolumn{2}{c}{Traffic} \\
            \hline
            Metrics & \multicolumn{1}{c}{F1$\uparrow$} & \multicolumn{1}{c|}{Cover$\uparrow$} & \multicolumn{1}{c}{F1$\uparrow$} & \multicolumn{1}{c|}{Cover$\uparrow$} & \multicolumn{1}{c}{F1$\uparrow$} & \multicolumn{1}{c}{Cover$\uparrow$} \\
            \hline
            {TS-CPP+H}   & 0.7003{\tiny $\pm$ 0.0029}  & 0.6444{\tiny $\pm$ 0.0148} & 0.7105{\tiny $\pm$ 0.0260} & 0.6720{\tiny $\pm$ 0.0320} & 0.3673{\tiny $\pm$ 0.0370} & 0.5315{\tiny $\pm$ 0.0313} \\
            {GDN+H}   & 0.6755{\tiny $\pm$ 0.0441}  & 0.6902{\tiny $\pm$ 0.0254} & 0.7181{\tiny $\pm$ 0.0319} & 0.7123{\tiny $\pm$ 0.0132} & 0.3473{\tiny $\pm$ 0.0067} & 0.5276{\tiny $\pm$ 0.0040 }   \\
            {OCGTL+H} & 0.7092{\tiny $\pm$ 0.0301}  & 0.7299{\tiny $\pm$ 0.0437} & 0.9248{\tiny $\pm$ 0.0207} & 0.8854{\tiny $\pm$ 0.0134} & 0.3674{\tiny $\pm$ 0.0379} & 0.5713{\tiny $\pm$ 0.0194}   \\
            sGNN+H & 0.7109{\tiny $\pm$ 0.0082}  & 0.7372{\tiny $\pm$ 0.0050} & 0.7061{\tiny $\pm$ 0.0019} & 0.6458{\tiny $\pm$ 0.0031} & 0.3611{\tiny $\pm$ 0.0257} & 0.5329{\tiny $\pm$ 0.0105}   \\
            \hline
            DETect & 0.4862{\tiny $\pm$ 0.0507} & 0.6559{\tiny $\pm$ 0.0284} & 0.2064{\tiny $\pm$ 0.0807}  & 0.4408{\tiny $\pm$ 0.0416}  & 0.3479{\tiny $\pm$ 0.0875}  & 0.5479{\tiny $\pm$ 0.0558}    \\
            \hline
            HSTCL  &\textbf{0.7757{\tiny $\pm$ 0.0026}}  &\textbf{0.7810{\tiny $\pm$ 0.0116 }}   &\textbf{0.9352{\tiny $\pm$ 0.0096 }}   &\textbf{0.9235{\tiny $\pm$ 0.0107 }}   &\textbf{0.3928{\tiny $\pm$ 0.0235 }}   &\textbf{0.5872{\tiny $\pm$ 0.0093 }}     \\
            \hline
            Improvement  &+59.54\%  &+19.07\% &+353.10\% &+109.51\% &+12.91\% &+7.17\%   \\
            \hline
            \end{tabular}%
    \label{tab:baselines}%
  \end{table*}%
The detecting performance of different methods is shown in Table~\ref{tab:baselines}. The metrics of DETect are relatively low on all datasets, showing that emergence detection can be difficult for traditional methods even on the simulation datasets. Deep learning methods generally outperform DETect in both metrics, and HSTCL achieves the highest performance. The results verify the superiority of our framework, which can capture emergence-related spatial patterns and model nonlinear spatio-temporal dynamics. sGNN+H is relatively poor on the Pedestrian dataset. It ignores the temporal dependence between adjacent graph snapshots and simply averages the similarities within the time window. HSTCL captures the temporal dependence via the STEs and learns a short-term representation of the system within the time window, which can better reflect the system-level variation in the latent space. GDN+H calculates the detecting score based on next-step prediction error, which can be sensitive to the noise in the states. HSTCL calculates the score based on the consistency of representations, which is resistant to potential noise. HSTCL jointly preserves spatial and temporal consistency, and thus outperforms TS-CPP+H which ignores spatial consistency and OCGTL+H which ignores temporal consistency.

\subsection{Ablation Study}
\begin{figure*}[!t]
    \centering
    \subfloat[States of road nets.]{\includegraphics[width=0.45\textwidth]{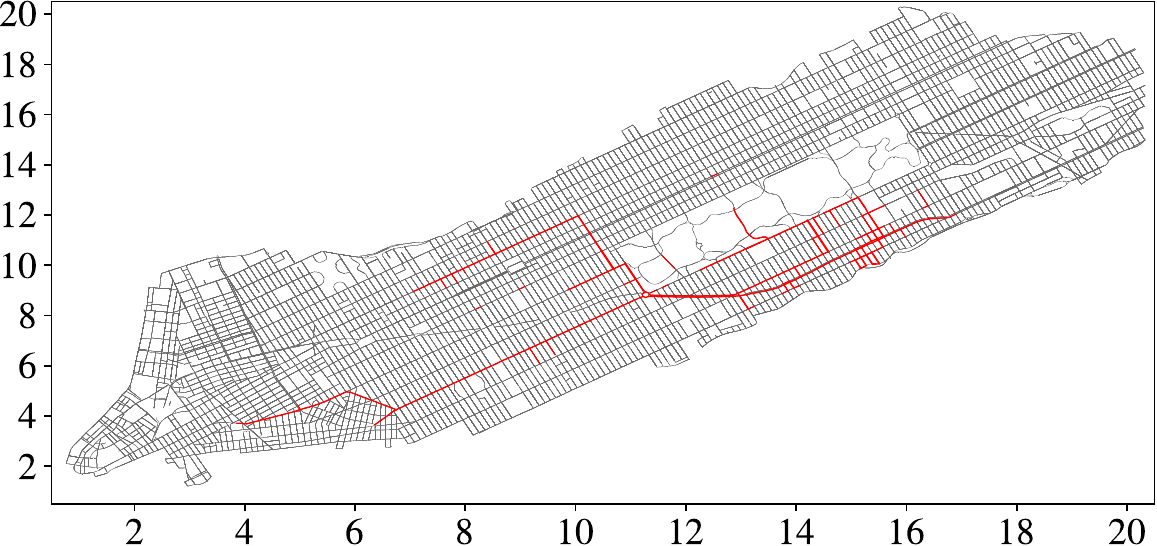}}
    \label{fig:road_met}
    \hfil
    \subfloat[States of regions.]{\includegraphics[width=0.45\textwidth]{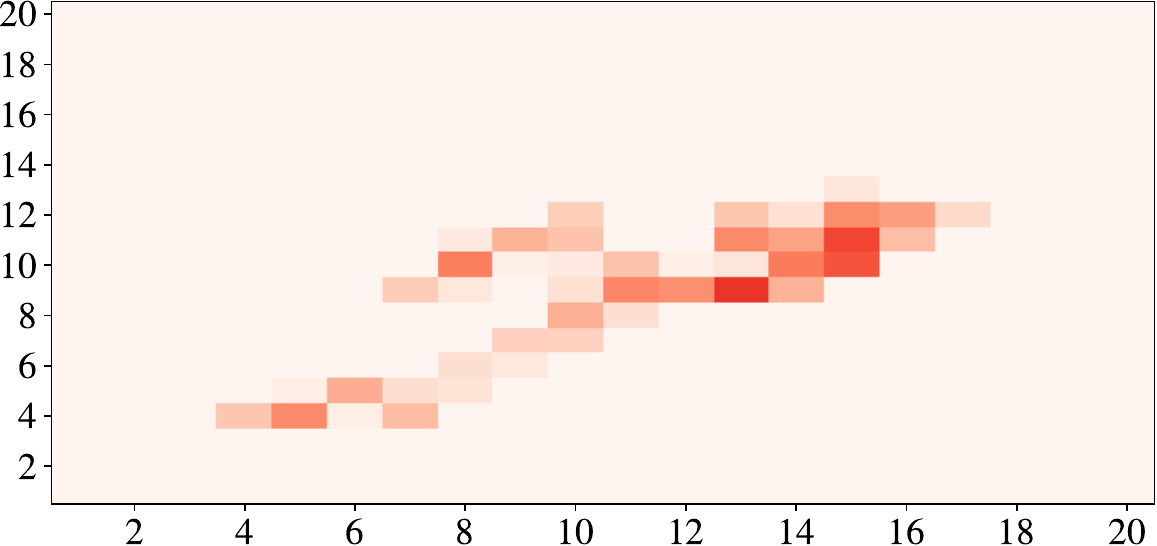}}
    \label{fig:region_aggregation}
    \caption{Case study of spatial patterns on the Traffic dataset. In (a), each line segment represents a road segment. Segments in red are congested, while those in gray are normal. In (b), each grid represents a region state. The darker the color, the more agent-level detections are collected. Best viewed in color.}
    \label{fig:spatial_patterns}
\end{figure*}
Some variants of HSTCL are introduced to verify the necessity of capturing emergence-related patterns and modeling agents' relationships with neighbors. HSTCL$_\text{Agent}$ removes system-level detection. HSTCL$_\text{Self}$ makes agent-level detection without modeling the spatial relationships, i.e., removing the spatial encoder and training without the spatial consistency loss. The Results are shown in Table~\ref{tab:spatiotemporal_consistency}.
\paragraph{Effect of System-Level Detection}
Without system-level detection, the average F1 score and covering metric of HSTCL$_\text{Agent}$ decrease by 0.0943 and  0.0970, respectively. The results verify that the spatial patterns of agent-level detecting results help with emergence detection. 

To see what patterns are captured by HSTCL, a case study is conducted on the Traffic dataset. Figure~\ref{fig:spatial_patterns} visualizes the congesting states of the road net and the region states when the network-level congestion forms. Figure~\ref{fig:spatial_patterns}(a) shows that congested road segments constitute a connected subnetwork with a diameter of 80, accounting for more than $\frac{1}{3}$ of the diameter of the road net. The phenomenon confirms the emergence of widespread congestion. Such emergence-related pattern is almost faithfully reflected in region states. The results also show that HSTCL can detect the emergence of widespread congestion even when the traffic flow is not provided.

\paragraph{Effect of Agent-Level Detection}
\begin{figure}[!t]
    \centering
    \includegraphics[width=0.95\columnwidth]{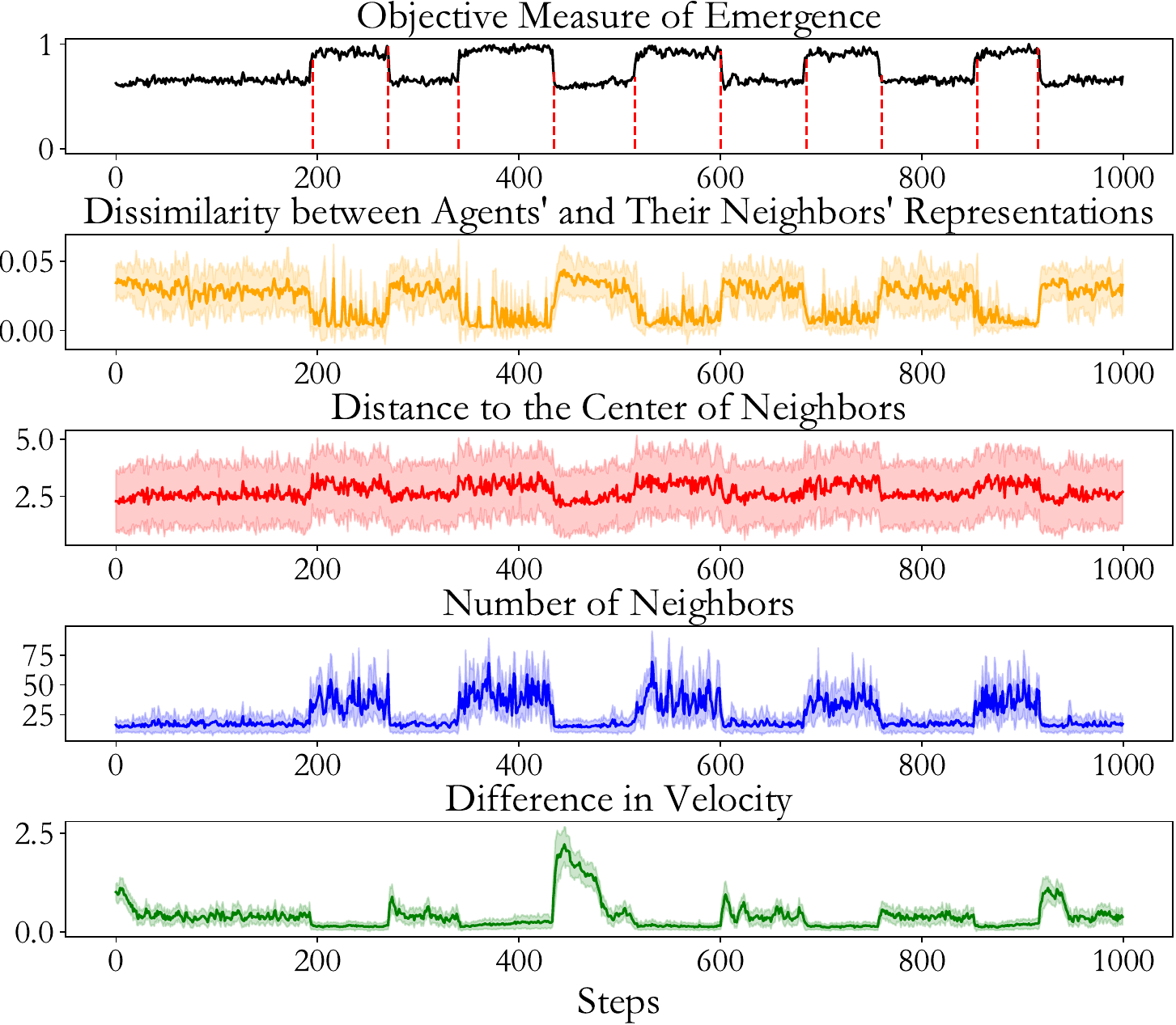}
    \caption{Case study of agents' relationships on the Flock dataset. The variation curves of the objective measure, agents' dissimilarity in latent space, and agents' relationships w.r.t. three indicators are visualized. Best viewed in color.}
    \label{fig:relationships}
\end{figure}
Compared with HSTCL$_\text{Agent}$, the average F1 score and the covering metric of HSTCL$_\text{Self}$ decrease by 0.0264 and 0.0171, respectively. The results show that modeling the nonlinear relationship between an agent and its neighbors is indispensable for agent-level detection. Note that HSTCL$_\text{Agent}$ is better than DETect on the Flock and Pedestrian datasets, and on par with it on the Traffic dataset, which again verifies the effectiveness of agent-level spatio-temporal modeling.

To see how HSTCL$_\text{Agent}$ captures the relationships, this paper conducts a case study on the Flock dataset. Inspired by \citet{otoole2017decentralised}, three indicators are used to measure the relationships w.r.t. agents' states, including the distance to the center of neighbors, the number of neighbors, and the difference between an agent's velocity and its neighbors' average velocity. The objective measure of emergence is set as a reference. Agents' and their neighbors' dissimilarity in representations stands for relationships in the latent space. The variation curves of the aforementioned metrics are shown in Figure~\ref{fig:relationships}. As expected, during the emergent period of flocking, birds get crowded, and thus, the distances are smaller, the neighbor count increases, and the difference in velocity decreases. Unexpectedly, the difference in velocity between 400 and 600 steps is larger than those in other periods. This anomalous phenomenon may be attributed to the flocking simulation rules. Birds are set to align during the emergent period and move randomly during the non-emergent period. The second emergent period is relatively long and thus the subsequent non-emergent period witnesses a sharp increase in velocity difference. The unstable behavior of the last metric shows that a single metric may not always be reliable for indicating emergence.

Different metrics present different patterns, yet these patterns are approximately captured by the latent representations. The tendency of the dissimilarity curve also agrees with that of the objective measure. These results show that our method can somehow comprehensively capture the relationships defined by some intuitive metrics w.r.t. agents' states.

\begin{table*}[!t]
    \renewcommand{\arraystretch}{1.3}
      \caption{Abalation study. Both the mean value and the standard deviation are reported. The best results of agent-level and system-level detection are in \textbf{bold}. The improvements are significant ($p$-value $<0.05$).}
    \centering
        \begin{tabular}{r|cc|cc|cc}
            \hline
            Datasets & \multicolumn{2}{c|}{Flock} & \multicolumn{2}{c|}{Pedestrian} & \multicolumn{2}{c}{Traffic} \\
            \hline
            Metrics & \multicolumn{1}{c}{F1$\uparrow$} & \multicolumn{1}{c|}{Cover$\uparrow$} & \multicolumn{1}{c}{F1$\uparrow$} & \multicolumn{1}{c|}{Cover$\uparrow$} & \multicolumn{1}{c}{F1$\uparrow$} & \multicolumn{1}{c}{Cover$\uparrow$} \\
            \hline
            HSTCL$_\text{Agent-S}$ & 0.6351{\tiny $\pm$ 0.0451 }  & \textbf{0.6962{\tiny $\pm$ 0.0438 }} & 0.7627{\tiny $\pm$ 0.0556} & 0.7430{\tiny $\pm$ 0.0361 } & 0.3189{\tiny $\pm$ 0.0265} & 0.5319{\tiny $\pm$ 0.0106}   \\
            HSTCL$_\text{Agent-T}$ & 0.6264{\tiny $\pm$ 0.0271 }  & 0.6847{\tiny $\pm$ 0.0275 } & 0.7773{\tiny $\pm$ 0.0119} & 0.7490{\tiny $\pm$ 0.0250 } & 0.3197{\tiny $\pm$ 0.0361} & 0.5227{\tiny $\pm$ 0.0164}   \\
            HSTCL$_\text{Agent}$ & \textbf{0.6705{\tiny $\pm$ 0.0114 }}  & \textbf{0.6960{\tiny $\pm$ 0.0168 }} & \textbf{0.7965{\tiny $\pm$ 0.0163}} & \textbf{0.7601{\tiny $\pm$ 0.0108 }} & \textbf{0.3538{\tiny $\pm$ 0.0027}} & \textbf{0.5445{\tiny $\pm$ 0.0090}}   \\
            \hline
            HSTCL$_\text{S}$  & 0.7155{\tiny $\pm$ 0.0317 }  & 0.7516{\tiny $\pm$ 0.0169 } & 0.9220{\tiny $\pm$ 0.0142} & 0.9089{\tiny $\pm$ 0.0199 } & 0.3690{\tiny $\pm$ 0.0266} & 0.5755{\tiny $\pm$ 0.0361}   \\
            HSTCL$_\text{T}$  & 0.7564{\tiny $\pm$ 0.0620 }  & 0.7680{\tiny $\pm$ 0.0310 } & 0.9278{\tiny $\pm$ 0.0122} & 0.9082{\tiny $\pm$ 0.0187 } & 0.3735{\tiny $\pm$ 0.0233} & \textbf{0.5880{\tiny $\pm$ 0.0219}}   \\
            
            HSTCL$_\text{Self}$ & 0.6413{\tiny $\pm$ 0.0303}  & 0.6899{\tiny $\pm$ 0.0319} & 0.7732{\tiny $\pm$ 0.0128} & 0.7211{\tiny $\pm$ 0.0325} & 0.3271{\tiny $\pm$ 0.0227} & 0.5382{\tiny $\pm$ 0.0088}  \\
            \hline
            HSTCL  &\textbf{0.7757{\tiny $\pm$ 0.0026}}  &\textbf{0.7810{\tiny $\pm$ 0.0116 }}   &\textbf{0.9352{\tiny $\pm$ 0.0096 }}   &\textbf{0.9235{\tiny $\pm$ 0.0107 }}   &\textbf{0.3928{\tiny $\pm$ 0.0235 }}   &\textbf{0.5872{\tiny $\pm$ 0.0093 }}     \\
            \hline
            \end{tabular}%
    \label{tab:spatiotemporal_consistency}%
  \end{table*}%
  \begin{figure*}[!t]
    \begin{minipage}{0.48\textwidth}
        \centering
        \includegraphics[width=\textwidth]{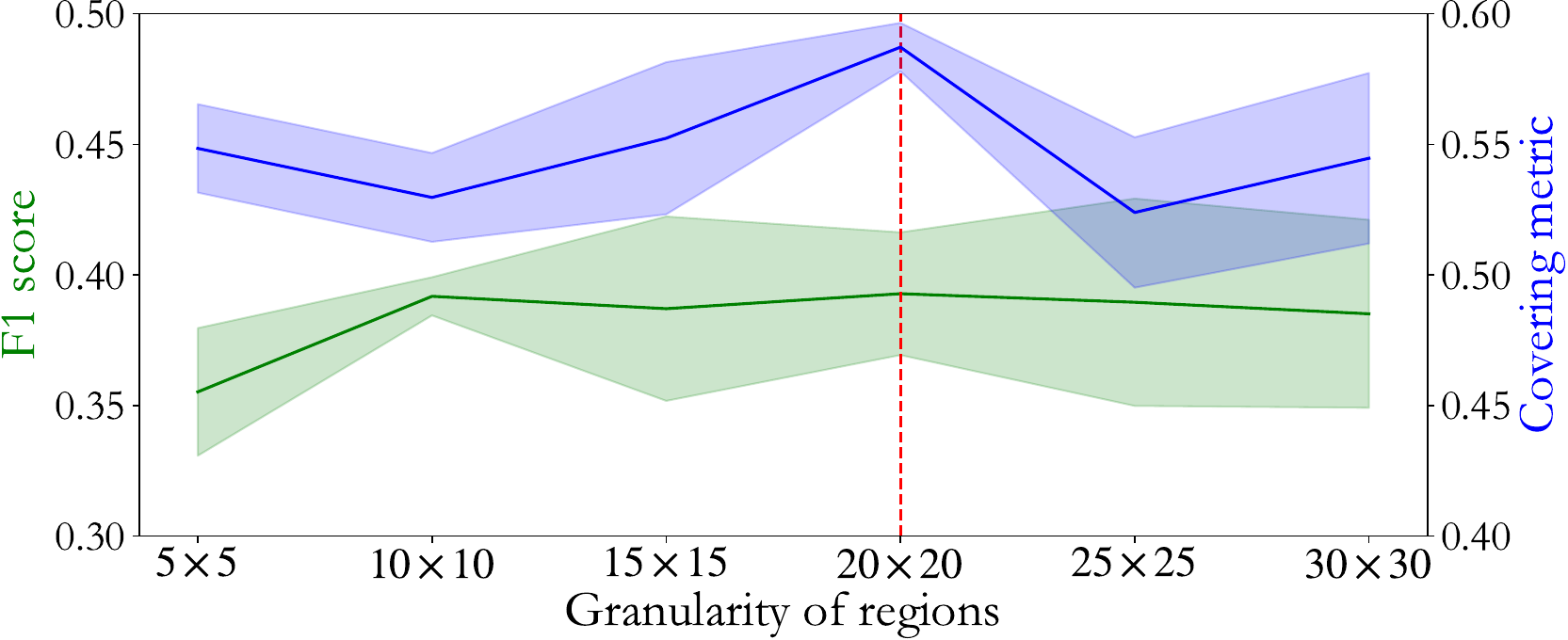}
        \caption{Effect of region granularity on the Traffic dataset. The dashed vertical line indicates the best results. Best viewed in color.}
        \label{fig:region_grain}
    \end{minipage}
    \hfil
    \begin{minipage}{0.48\textwidth}
        \centering
        \includegraphics[width=\textwidth]{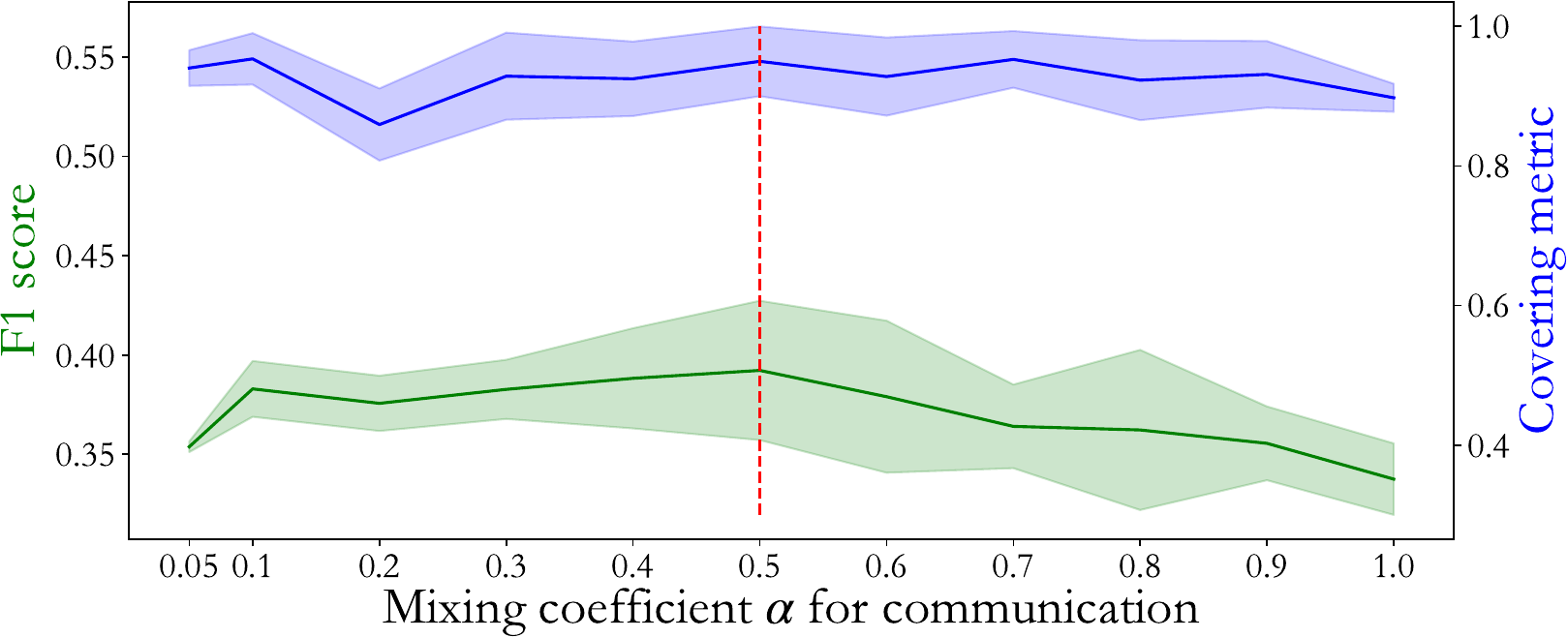}
        \caption{Effect of agents' communication on the Traffic dataset. The dashed vertical line indicates the best results. Best viewed in color.}
        \label{fig:communication}
    \end{minipage}
\end{figure*}
\paragraph{Effect of Spatial and Temporal Consistency Losses}
To validate the effectiveness of STCL for both agent-level and system-level detection, this paper introduces variants of HSTCL$_\text{Agent}$ and HSTCL trained with only the spatial or the temporal consistency loss. The resulting methods are denoted as HSTCL$_\text{Agent-S}$, HSTCL$_\text{Agent-T}$, HSTCL$_\text{S}$, and HSTCL$_\text{T}$, respectively. As shown in Table~\ref{tab:spatiotemporal_consistency}, removing any term in the loss function will lead to degenerated performance in most cases. For example, HSTCL$_\text{Agent-T}$ removes the agent-level spatial consistency loss, and noticeable drops in both metrics can be observed on all datasets. The results verify that inconsistency in spatial relation is an accurate indicator of emergence. Similarly, HSTCL$_\text{S}$ removes the system-level temporal consistency loss and both metrics decrease. The results verify that temporal inconsistency of the whole system helps to detect the emergent behavior. Thus, the spatial consistency loss and temporal consistency loss are complementary to learning discriminative representations for emergence detection.

\subsection{Hyperparameter Analysis}\label{sec:exp:hyperparameter}
\paragraph{Effect of Region Granularity} 
The area where agents move is split into many regions for system-level detection. The granularity of regions decides how many details are preserved for global analysis. To study the effect of region granularity, this paper trains the system-level detector on the Traffic dataset under several $N\times N$ grids, with $N\in\{5, 10, 15, 20, 25, 30\}$. The results are shown in Figure~\ref{fig:region_grain}. The F1 score is lowest when $N=5$. Maybe coarsening too much will result in inadequate information that cannot support accurate detection. The F1 score and covering metric increase on the whole as $N$ grows but start to decrease at $N=20$. An exception is that the covering metric decreases at $N=10$. The F1 score is the harmonic mean of the precision and the recall rate. Since the threshold $c$ for determining a change point is searched by maximizing the F1 score on the validation set, it is possible that the precision or the recall rate increases while the other metric decreases, leading to the growth of the F1 score and the drop of the covering metric. Some examples are provided in \textit{Appendix D}. When $N$ is too large, the regions are too small to collect sufficient feedback from agents and present stable spatial patterns. Thus, our method achieves the highest performance for $N=20$ with a moderate computational cost.

\paragraph{Effect of Mixing Coefficient for Communication} Agents communicate with neighbors to make agent-level detections. The mixing coefficient $\alpha$ in Eq.~\eqref{eq:communication} balances the importance of an agent's current observation and its neighbors' detecting scores. $\alpha$ is set to $0.05$ to keep consistent with DETect. To see how $\alpha$ affects the detecting accuracy, this paper evaluates HSTCL$_\text{Agent}$ on the Traffic dataset with $\alpha$ ranging from $0.1$ to $1$. As shown in Figure~\ref{fig:communication}, the F1 score and covering metric can be promoted by increasing $\alpha$, i.e., assigning a larger weight to agents' current observation. However, when $\alpha=1$, i.e., agents ignore the detecting scores of their neighbors, the F1 score drops significantly, verifying the necessity of communication. The results show the possibility of tuning $\alpha$ to improve the detecting accuracy. Furthermore, $\alpha$ can be personalized and adaptive over time. Optimizing the choices of $\alpha$ is left for future work.
\begin{figure}[!t]
    \centering
    \includegraphics[width=0.48\textwidth]{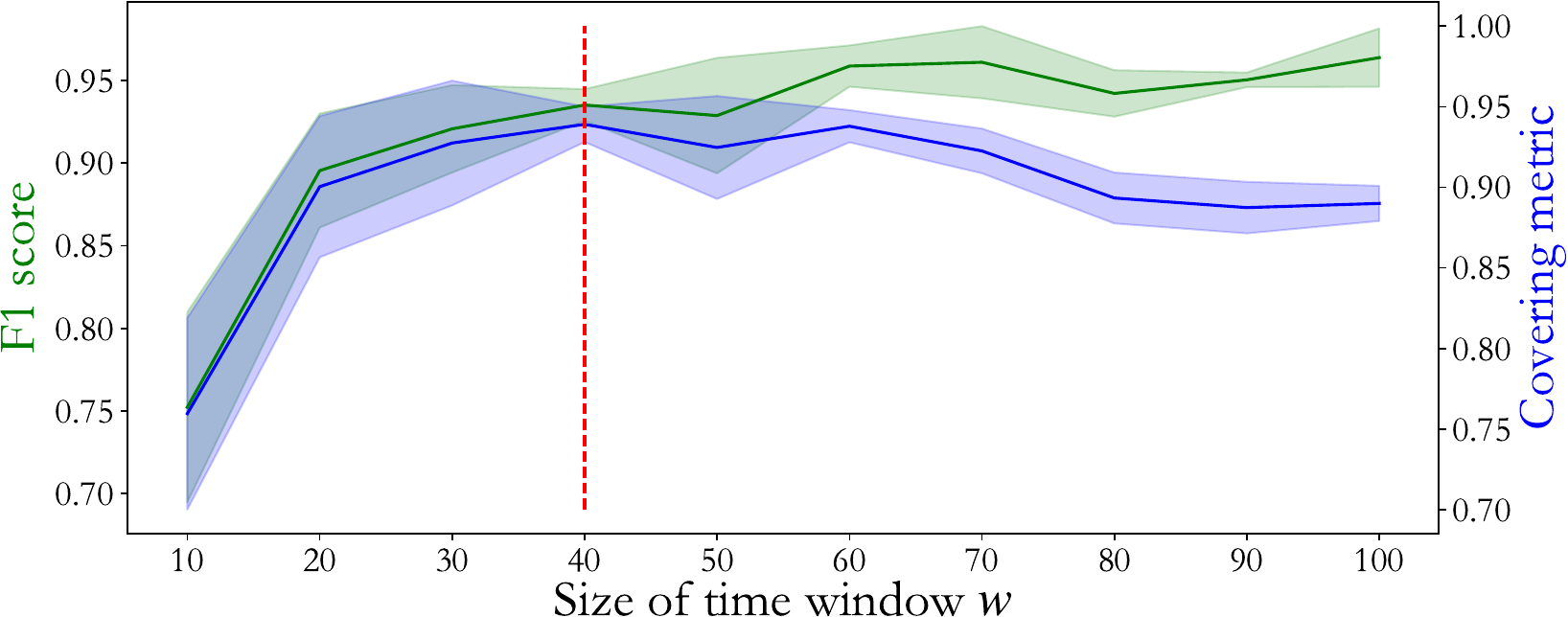}
    \caption{Effect of window size in system-level detection on the Pedestrian dataset. Best viewed in color.}
    \label{fig:duration}
\end{figure}
\begin{figure}[!t]
    \centering
    \includegraphics[width=0.48\textwidth]{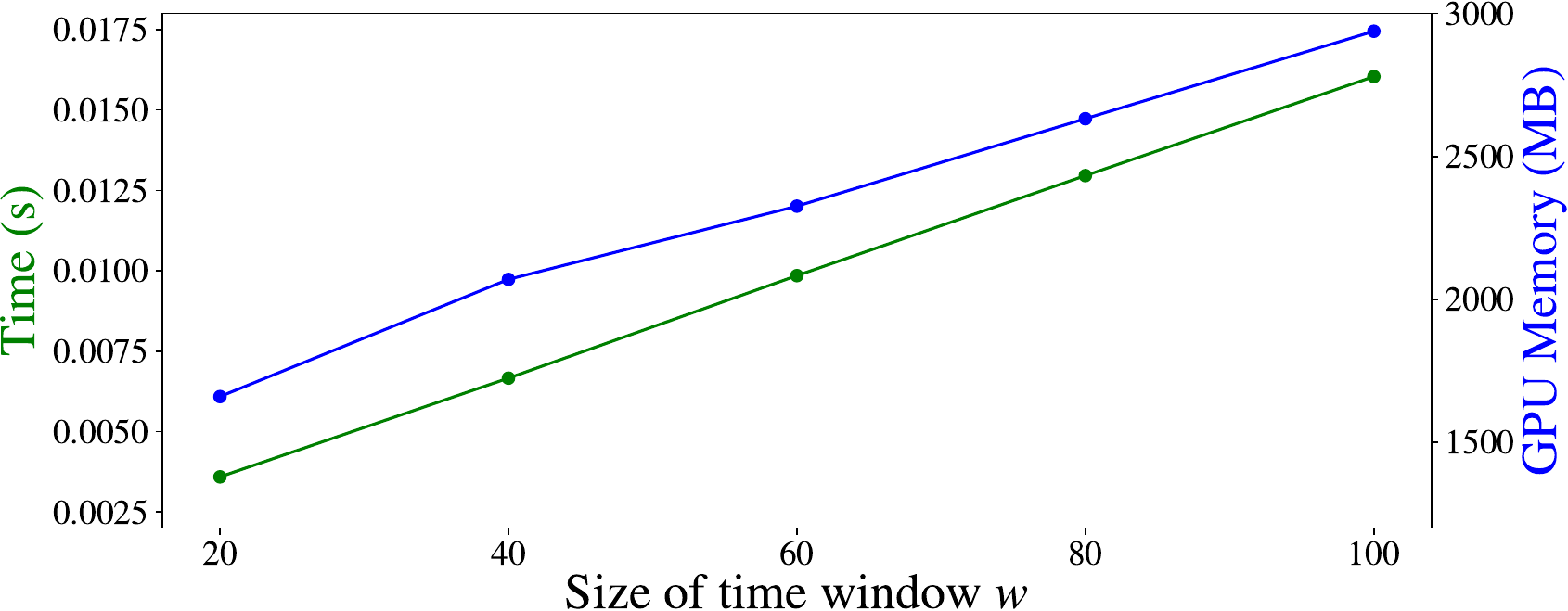}
    \caption{Running time and GPU memory consumption of system-level detection w.r.t. $w$ on the Pedestrian dataset.}
    \label{fig:running_time_memory_duration}
\end{figure}
\paragraph{Effect of Window Size for Emergence Detection}\label{app:duration}
The window size $w$ is the temporal scope of emergence detection. Like the granularity of regions, there is a tradeoff between precision and efficacy w.r.t. $w$. It is set to $10$ for agent-level detection, since the objective measure of emergence is evaluated every 50 steps and the states of agents are downsampled every 5 steps. In this way, agents can make relatively accurate and timely detections. However, performance degeneration is observed for system-level detection with the same window size. It is conjectured that system-level detection needs a larger temporal scope with coarse-grained spatial information. The effect of $w$ is studied on the Pedestrian dataset, with $w\in\{10, 20,\dots,100\}$. As shown in Figure~\ref{fig:duration}, the F1 score generally increases as $w$ grows, while the covering metric peaks at $w=40$. Although the F1 score peaks at $w=60$, both the running time and the memory consumption will increase, as depicted in Figure~\ref{fig:running_time_memory_duration}. Since online detection is sensitive to the computational cost, $w$ is set to 40 for system-level detection to achieve a good trade-off between accuracy and efficiency. This choice is also practical since the global monitor generally has a larger capacity than a single agent.

\paragraph{Effect of Threshold $\theta$}
\begin{figure}[!t]
    \centering
    \includegraphics[width=0.48\textwidth]{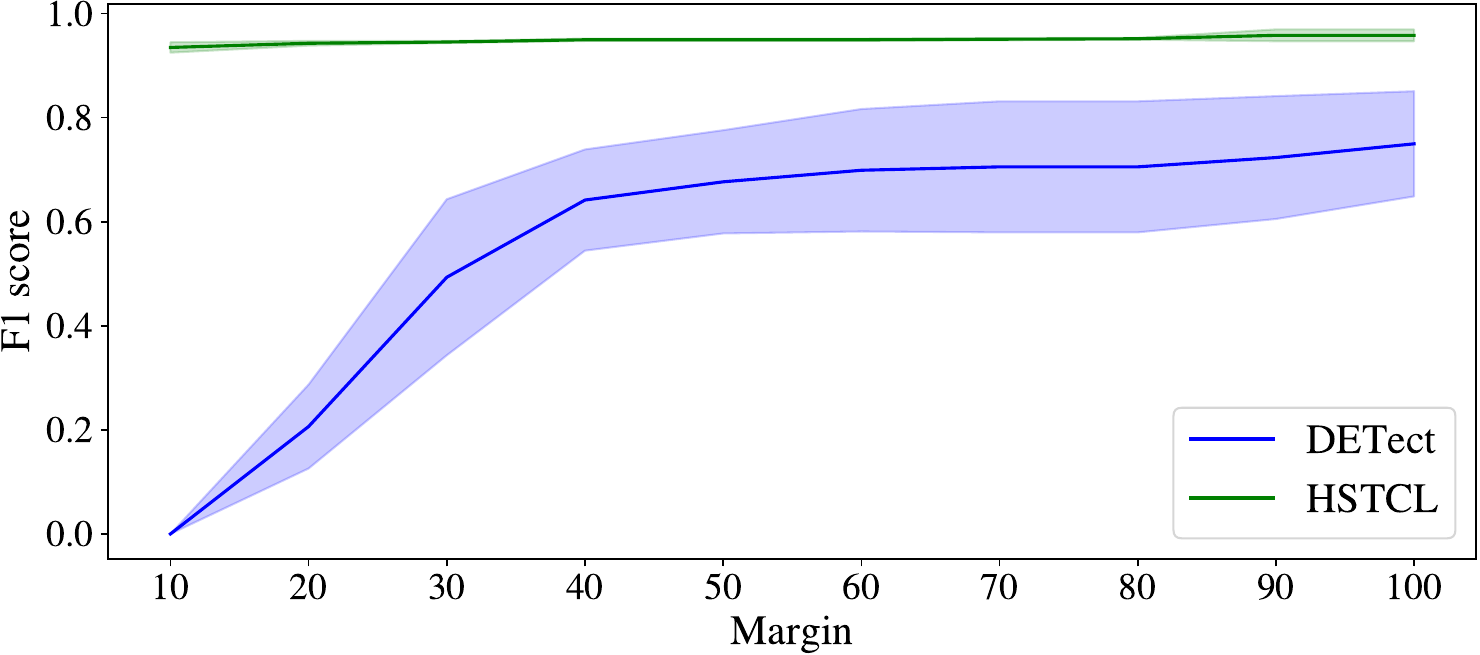}
    \caption{Effect of threshold $\theta$ on the Pedestrian dataset. Best viewed in color.}
    \label{fig:threshold}
\end{figure}
In the experiments, the threshold $\theta$ is set to $20$ on all datasets for fairly comparing the detecting precision of different methods. Apparently, the F1 scores are affected by the choices of $\theta$. This paper evaluates the F1 scores of HSTCL and DETect on the Pedestrian dataset for $\theta\in\left\{10,20,\dots,100\right\}$, and the results are shown in Figure~\ref{fig:threshold}. For both methods, the F1 score grows approximately as $\theta$ increases, because a larger tolerance allows to include more detected change points. On the whole, HSTCL consistently achieves higher detecting precision than DETect for all $\theta$, yet the difference narrows down as $\theta$ increases. Besides, the variance of DETect's F1 scores tends to grow up for a larger $\theta$, while HSTCL preserves a considerably smaller variance, showing that the performance of HSTCL is more stable. In practice, a relatively smaller threshold is more favorable, because detected change points with smaller displacement help to detect the emergence more timely.
\begin{figure}[!t]
    \centering
    \includegraphics[width=0.48\textwidth]{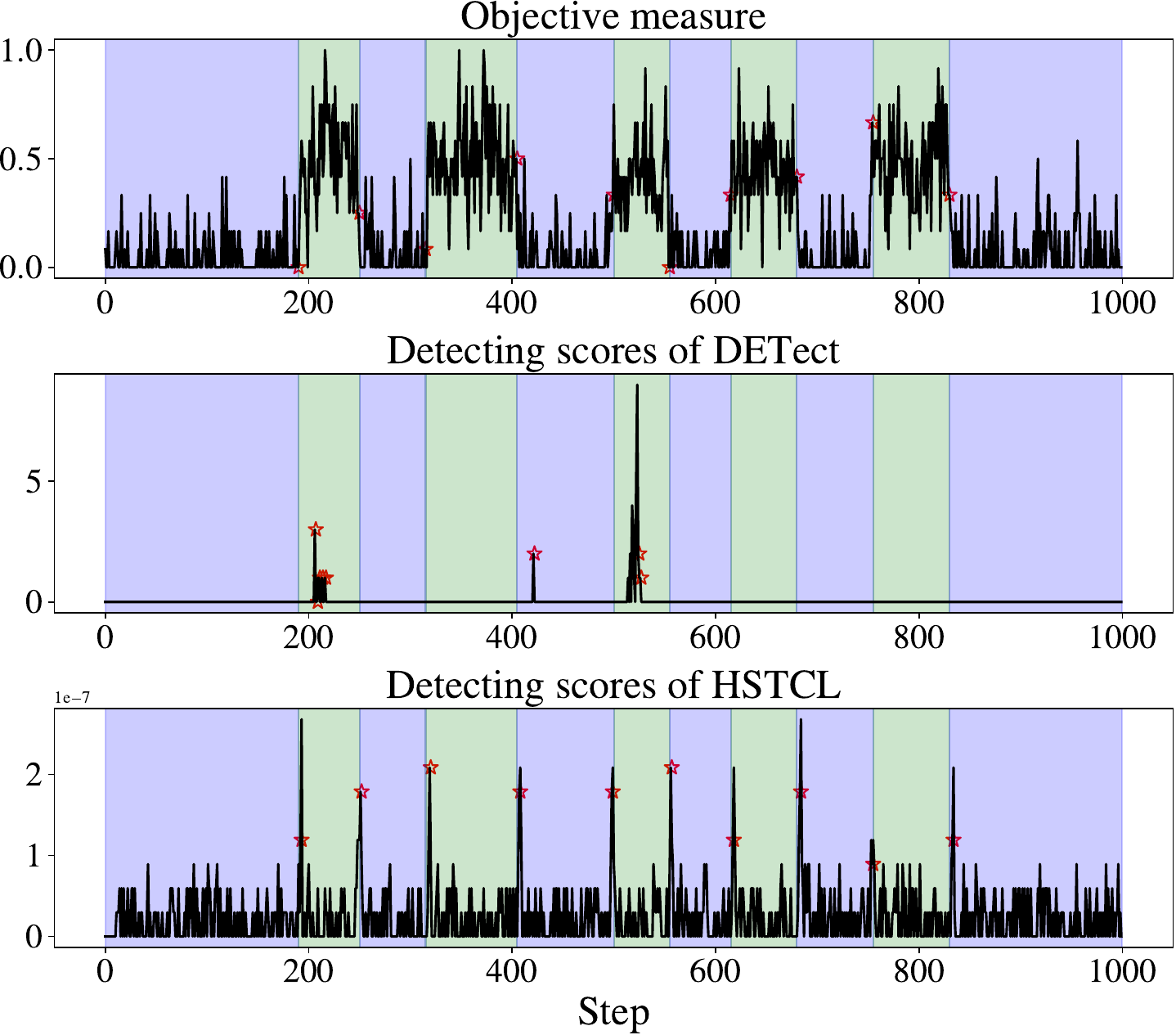}
    \caption{Visualization of the ground truth change points (\textit{top}), and change points detected by DETect (\textit{middle}) and HSTCL (\textit{bottom}) on the Pedestrian dataset. The variation curves are in black. The change points are marked as stars. The periods of emergence are in green. The normal periods are in purple. Best viewed in color.}
    \label{fig:change_points}
\end{figure}

To further differentiate the detecting quality of DETect and HSTCL, this paper visualizes their detected change points on the Pedestrian dataset. As shown in Figure~\ref{fig:change_points}, DETect fails to detect change points in several periods, and the detected points are relatively far from the nearest ground truth. By contrast, HSTCL successfully detects all change points with significantly smaller deviations.

\paragraph{Sensitivity Analysis} It is crucial to analyze the sensitivity of HSTCL w.r.t. to the initial parameters for instructing its real-world applications. As reported in Table~\ref{tab:baselines} and Table~\ref{tab:spatiotemporal_consistency}, the standard deviations for both agent-level and system-level detection of HSTCL are relatively small, which indicates that HSTCL is relatively robust to the stochastic nature of deep learning, including the weight initialization and the training process. In Section~\ref{sec:exp:hyperparameter} (a)-(d), this paper analyzes the effect of region granularity $N\times N$, mixing coefficient $\alpha$ for communication, window size $w$, and the threshold $\theta$ for evaluating metrics. To summarize, HSTCL is sensitive to $N\times N$, $\alpha$ and $w$, but is relatively stable w.r.t. $\theta$. In practice, one can use the historical data as a validation set to determine essential parameters and adjust these parameters when there is distribution shift of the online data. Specifically, $\alpha$ can be tuned without retraining the agent-level model. Adjusting $N$ may not affect the deployment of edge monitors because one monitor can collect agents' feedback from several regions, but the system-level model should be updated. Adjusting $w$ requires to update the agent-level model and the system-level model sequentially. When the system is nonstationary, it can be analyzed on multiple time scales simultaneously to achieve timely detection with relatively low cost~\cite{wang2021multiscale}.

\subsection{Running Time Analysis}\label{sec:exp:running_time}
\begin{figure}[!t]
    \centering
    \includegraphics[width=0.47\textwidth]{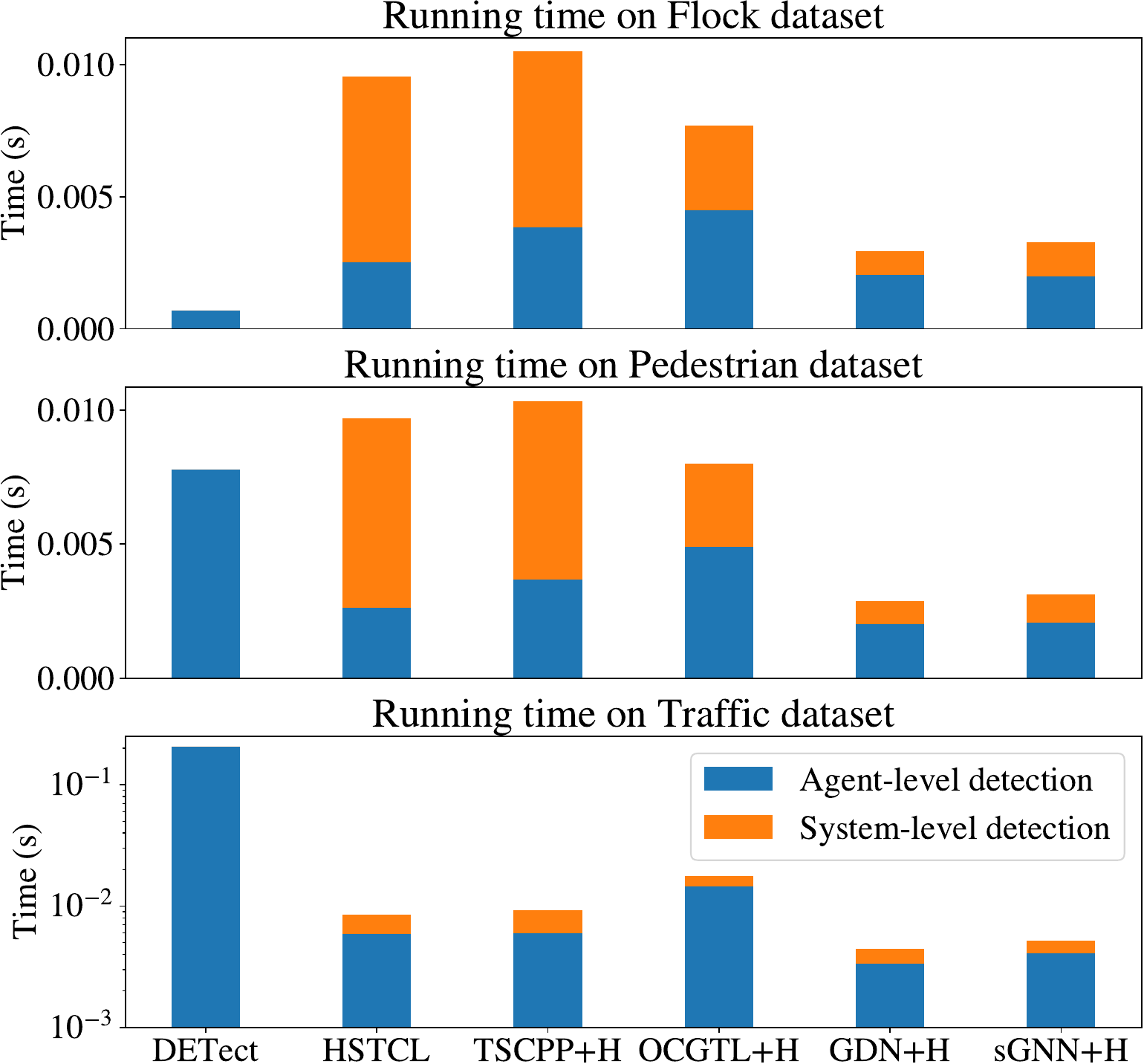}
    \caption{Runtime of different methods on all datasets. Best viewed in color.}
    \label{fig:running_time}
\end{figure}

To study the efficiency of all methods for online emergence detection, this paper reports their average running time per step for agent-level detection and system-level detection in Figure~\ref{fig:running_time}. Although DETect is the most efficient on the Flock dataset which contains only 150 agents, its running time grows steeply w.r.t. the number of agents and peaks on the Traffic dataset which contains 2,522 agents. Such inefficiency may be due to its concrete implementation in NetLogo. Among deep learning methods within our framework, GDN+H and sGNN+H are faster than others because of their simplicity in spatio-temporal modeling. The overall running time of HSTCL, TSCPP+H, and OCGTL+H are comparable. However, OCGTL+H takes more time in agent-level detection than other methods, which may be due to its computationally costly design of transformation learning. Notably, on the large-scale Traffic dataset, the efficiency of HSTCL becomes closer to sGNN+H, which demonstrates the scalability of HSTCL.

On the Traffic dataset where agents are significantly more than regions, the overall running time is dominated by agent-level detection. On the Flock and Pedestrian datasets where regions are more than agents, system-level detection dominates the running time for HSTCL and TSCPP+H. For GDN+H, OCGTL+H, and sGNN+H, agent-level detection takes more time. This may be attributed to the simplicity or ignorance of temporal modeling, and that agent graphs are denser than region graphs. Note that the running time of all methods are evaluated on a single machine. In practice, agent-level detection can be accomplished by each agent in parallel. Thus, it is expected that all distributed detection methods scale well as the number of agents grows.

\section{Conclusion}
This paper proposes a hierarchical framework named HSTCL for emergence detection in CAS under the distributed setting. By aggregating agent-level detecting results from bottom-up, HSTCL learns a system representation that captures emergence-related patterns. Nonlinear relationships between agents and their neighbors are encoded in agent representations through STE. These representations are learned in a self-supervised manner by preserving the spatio-temporal consistency. HSTCL surpasses the traditional methods and deep learning methods on three datasets with well-known yet hard-to-detect emergent phenomena. HSTCL is flexible to incorporate deep learning methods from graph-level CPD and anomaly detection for effective emergence detection.

In the future, HSTCL can be extended from three dimensions: data, problems and methods. On the data dimension, we plan to acquire real-time data streams from devices in internet of things to test HSTCL's effectiveness in live environments. Besides, allowing addition and removal of agents will make the simulators more realistic and brings extra challenges to dynamic graph learning~\cite{skarding2021foundations}
. It is also promising to utilize simulators based on large language models~\cite{gao2023s} to promote the quality of simulation data and make the evaluation results more convincing. These simulators can better incorporate domain knowledge, and recover complex agent behaviors beyond predefined simulation rules. When real-world observations of agent states are sparse and incomplete, graph learning methods can be developed to complement missing information
and possible auxiliary information can be leveraged to assist the detection~\cite{zhang2019onlinecpd}. When the systems contains heterogeneous agents, category-aware STEs and learning strategies can be designed to modeling heterogeneous dynamics and interacting patterns~\cite{chen2024heterogeneous}. On the problem dimension, it is more realistic to consider partially missing messages and time delays in agents' communication, which will trigger the research of more robust detectors. On the method dimension, distributed GNNs and advanced CPD methods can further boost the performance of emergence detection. Besides, it is promising to develop graph learning methods that can capture emergence-related higher-order structures of a system.

\bibliographystyle{IEEEtranN}
\bibliography{IEEEabrv,ref}

\includepdf[pages=1-8]{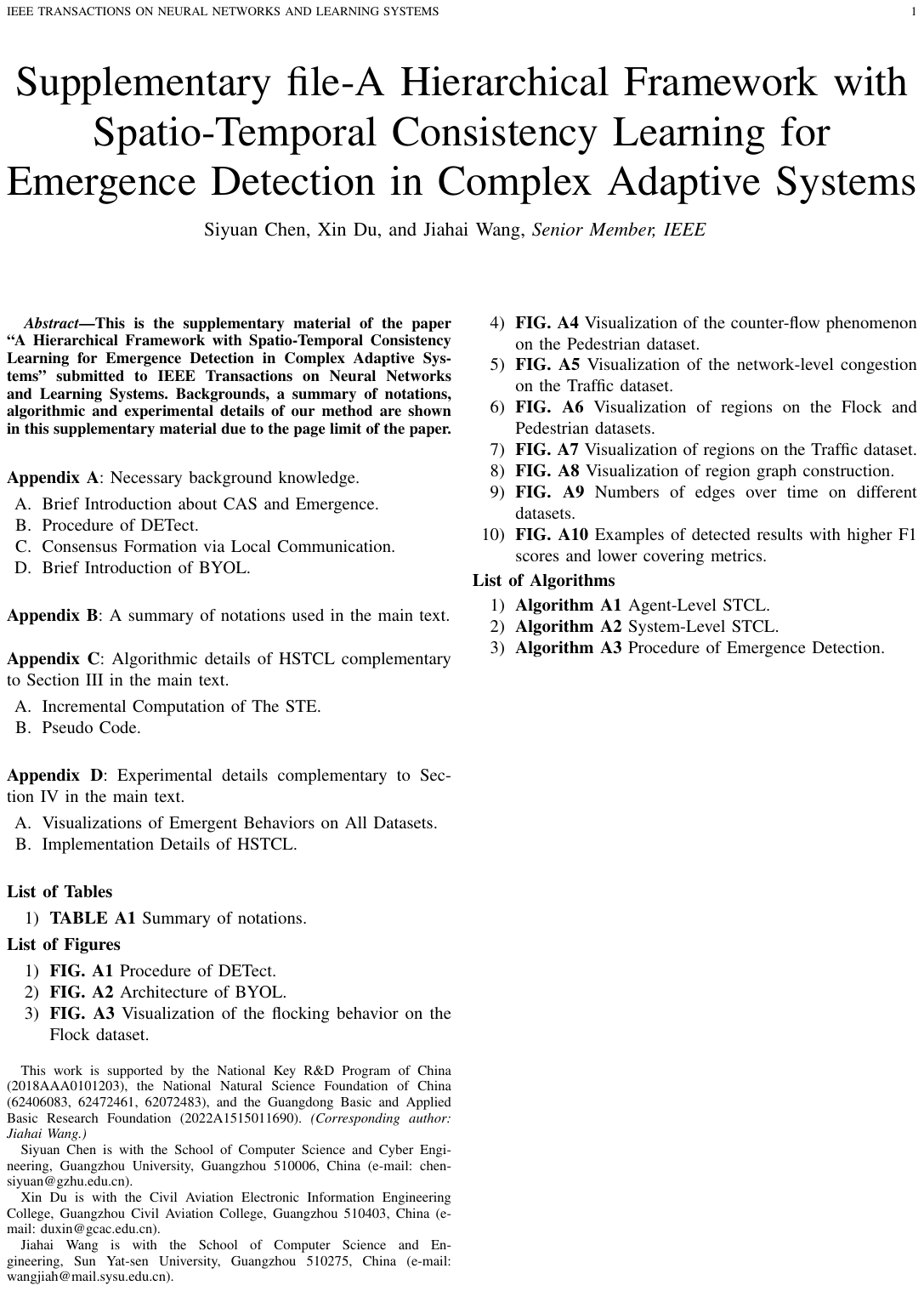}
\end{document}


\title{Supplementary file-A Hierarchical Framework with Spatio-Temporal Consistency Learning for Emergence Detection in Complex Adaptive Systems}
\author{Siyuan Chen, Xin Du, and Jiahai Wang,~\IEEEmembership{Senior Member,~IEEE}
\thanks{
    This work is supported by the National Key R\&D Program of China (2018AAA0101203), the National Natural Science Foundation of China (62406083, 62472461, 62072483), and the Guangdong Basic and Applied Basic Research Foundation (2022A1515011690). \textit{(Corresponding author: Jiahai Wang.)} 
  }
\thanks{
  Siyuan Chen is with the School of Computer Science and Cyber Engineering, Guangzhou University, Guangzhou 510006, China (e-mail: chensiyuan@gzhu.edu.cn).
  
  Xin Du is with the Civil Aviation Electronic Information Engineering College, Guangzhou Civil Aviation College, Guangzhou 510403, China (e-mail: duxin@gcac.edu.cn).
  
  Jiahai Wang is with the School of Computer Science and Engineering, Sun Yat-sen University, Guangzhou 510275, China (e-mail: wangjiah@mail.sysu.edu.cn).
  }
}

\markboth{IEEE Transactions on Neural Networks and Learning Systems}%
{Shell \MakeLowercase{\textit{et al.}}: A Sample Article Using IEEEtran.cls for IEEE Journals}


\maketitle

\begin{abstract}
  This is the supplementary material of the paper ``A Hierarchical Framework with Spatio-Temporal Consistency Learning for Emergence Detection in Complex Adaptive Systems'' submitted to IEEE Transactions on Neural Networks and Learning Systems. Backgrounds, a summary of notations, algorithmic and experimental details of our method are shown in this supplementary material due to the page limit of the paper.
\end{abstract}

\noindent\textbf{Appendix~A}: Necessary background knowledge.
\begin{itemize}
  \item [A.] Brief Introduction about CAS and Emergence.
  \item [B.] Procedure of DETect.
  \item [C.] Consensus Formation via Local Communication.
  \item [D.] Brief Introduction of BYOL.
\end{itemize}

~\\
\noindent\textbf{Appendix~B}: A summary of notations used in the main text.

~\\
\noindent\textbf{Appendix~C}: Algorithmic details of HSTCL complementary to Section~III in the main text.
\begin{itemize}
  \item [A.] Incremental Computation of The STE.
  \item [B.] Pseudo Code.
\end{itemize}

~\\
\noindent\textbf{Appendix D}: Experimental details complementary to Section~IV in the main text.
\begin{itemize}
  \item [A.] Visualizations of Emergent Behaviors on All Datasets.
  \item [B.] Implementation Details of HSTCL.
\end{itemize}

~\\
\noindent\textbf{List of Tables}

\begin{enumerate}
  \item \textbf{TABLE A1} Summary of notations.
\end{enumerate}

\noindent\textbf{List of Figures}
\begin{enumerate}
  \item \textbf{FIG. A1} Procedure of DETect.
  \item \textbf{FIG. A2} Architecture of BYOL.
  \item \textbf{FIG. A3} Visualization of the flocking behavior on the Flock dataset.
  \item \textbf{FIG. A4} Visualization of the counter-flow phenomenon on the Pedestrian dataset.
  \item \textbf{FIG. A5} Visualization of the network-level congestion on the Traffic dataset.
  \item \textbf{FIG. A6} Visualization of regions on the Flock and Pedestrian datasets.
  \item \textbf{FIG. A7} Visualization of regions on the Traffic dataset.
  \item \textbf{FIG. A8} Visualization of region graph construction.
  \item \textbf{FIG. A9} Numbers of edges over time on different datasets.
  \item \textbf{FIG. A10} Examples of detected results with higher F1 scores and lower covering metrics.
\end{enumerate}

\noindent\textbf{List of Algorithms}
\begin{enumerate}
  \item \textbf{Algorithm A1} Agent-Level STCL.
  \item \textbf{Algorithm A2} System-Level STCL.
  \item \textbf{Algorithm A3} Procedure of Emergence Detection.
\end{enumerate}

\clearpage
\appendices
\renewcommand\thefigure{A\arabic{figure}}
\renewcommand\thetable{A\arabic{table}}
\renewcommand\theequation{A\arabic{equation}}
\setcounter{page}{1}
\setcounter{figure}{0}
\setcounter{table}{0}
\setcounter{equation}{0}
\setcounter{footnote}{0}
\setcounter{theorem}{0}
\setcounter{proposition}{0}
\setcounter{corollary}{0}
\section{Backgrounds}
\subsection{Brief Introduction about CAS and Emergence}
CAS are dynamic systems where individual and collective behaviors can adapt to changing interactions and environments~\cite{miller2009complex,Carmichael2019complex}. CAS present various intrinsic macro-level properties like complexity, self-organization, self-similarity, and emergence, which make them distinct from multi-agent systems that put more emphasis on micro-level and meso-level properties~\cite{miller2009complex,Carmichael2019complex}. This paper focuses on the intriguing emergent property, which is applicable to study important real-world phenomena like emerging topics in social networks~\cite{takahashi2014discovering}, synchronization on complex networks~\cite{zhu2023adaptive} and phase transition on traffic networks~\cite{li2015percolation}.

Specifically, the weak emergence that can be deduced through simulations is studied in this paper~\cite{artime2022origin}. It is more scientifically relevant, and the performance of methods can be evaluated empirically. The strong emergence, e.g., life that emerges from genes and consciousness that emerges from neurons are out of our scope. It can not be deduced even in principle. The research on strong emergence can be substantially more difficult.

The research of emergence is quite interdisciplinary, spreading over complexity science, social science, computer science, etc. Research topics of emergence include its quantitative definition, detection, prediction, control, etc. This paper is devoted to emergence detection, since its prediction and control is hard by nature~\cite{kalantari2020emergence}. Traditional methods apply agent-based modeling~\cite{niazi2011agent} to study emergence. By contrast, this paper develops data-driven spatio-temporal learning techniques for emergence detection, which may benefit the research on other aspects of emergence.

\subsection{Procedure of DETect}
\begin{figure*}[!b]
    \centering
    \includegraphics[width=0.7\textwidth]{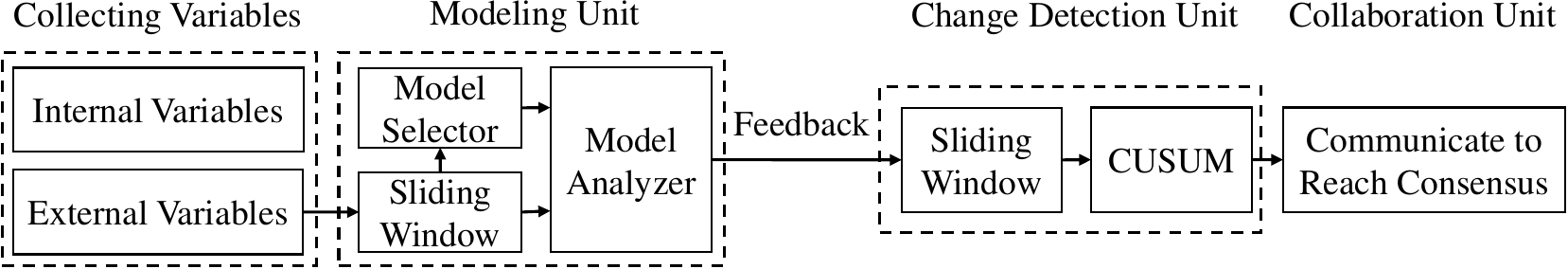}
    \caption{Procedure of DETect.}
    \label{fig:DETect}
\end{figure*}

DETect~\cite{otoole2017decentralised} is a decentralized method that utilizes agents' local observation and collaboration among agents to achieve online emergence detection. Its procedure is shown in Figure~\ref{fig:DETect}.

DETect is composed of three units, the modeling unit, the change detection unit, and the collaboration unit. They are responsible for modeling the relationship between an agent and its neighbors, detecting if the relationship changes significantly, and communicating with neighbors to reach a consensus on the formation or evaporation of emergence. Before detecting emergence, each agent records its own state and its neighbors' states within a time window, called the internal variables and the external variables, respectively. For example, in urban traffic systems, each car records its speed and heading direction as interval variables, and records neighbors' average heading direction, average speed, the distance to its nearest neighbor, and the number of neighbors as external variables. The relation between internal variables and external variables reflects the relationship between an agent and its neighbors. The three units of DETect are described as follows.

\paragraph{Modeling Unit} This unit aims to model the relationship between internal variables and external variables via linear regression. The statistical significance test is applied to each pair of internal variable and external variable at each time step. The $p$-value indicates the strength of the relation. A smaller value means a stronger relation.

\paragraph{Change Detection Unit} This unit uses the $p$-values output by the modeling unit to make change-point detection. The CUMSUM~\cite{page1954continuous} algorithm based on cumulated change is applied to detect change points in the $p$-value sequence w.r.t each pair of internal variable and external variable. Once the relation between any pair of variables changes noticeably, the relationship between an agent and its neighbors is thought to change significantly.

\paragraph{Collaboration Unit} This unit allows agents to share with neighbors their belief of emergence, a scalar ranging in $[0, 1]$, to reach a consensus on emergence. At each time step, each agent selects a random neighbor to communicate, and updates its belief of emergence as the average value between itself and the neighbor. The convex combination of the belief and a binary variable indicating the change point an agent detects serves as the final belief of emergence.

An agent sends feedback when its belief of emergence exceeds some given threshold. When the number of feedback grows significantly larger, DETect confirms the formation or evaporation of emergence. It can be seen that DETect requires a global monitor the collect all agents' feedback. Therefore, it is more appropriate to regard DETect as a distributed method with a weak center rather than a fully decentralized method.

\subsection{Consensus Formation via Local Communication}
This subsection analyzes the conditions of consensus formation among agents via local communication. Recall from the main text that the update of agent-level detecting scores via communication is formulated as
\begin{equation}
    s^{\tau+1}_j = \alpha \cdot d\left(\mathbf{h}^{(\tau)}_j,\mathbf{h}^{(\tau-1)}_j\right) + \frac{(1-\alpha)}{|\mathcal{N}_j^\tau|+1}\sum_{i\in\mathcal{N}^\tau_j\cup\{j\}}s^\tau_i.
    \label{eq:communication}
\end{equation}

Eq.~\eqref{eq:communication} can be rewritten in a matrix form. Denote the score vector and the dissimilarity vector as
\begin{equation}
    \begin{aligned}
    \mathbf{s}^{\tau}&=\left(s^\tau_1,\dots,s^\tau_{|\mathcal{V}|}\right)^\top,\\
    \mathbf{d}^{\tau}&=\left(d\left(\mathbf{h}^{(\tau)}_1,\mathbf{h}^{(\tau-1)}_1\right),\dots,d\left(\mathbf{h}^{(\tau)}_{|\mathcal{V}|},\mathbf{h}^{(\tau-1)}_{|\mathcal{V}|}\right)\right)^\top.
    \end{aligned}
\end{equation}
Let $\mathbf{A}^\tau\in\{0, 1\}^{|\mathcal{V}|\times|\mathcal{V}|}$ be the adjacency matrix of the communication graph at time $\tau$, $\mathbf{A}^\tau+\mathbf{I}$ be the adjacency matrix of the graph with self-loops, and $\mathbf{D}^\tau$ be the degree matrix of $\mathbf{A}^\tau+\mathbf{I}$. The row-normalized random walk matrix is denoted as $\widetilde{\mathbf{A}}^\tau=(\mathbf{D}^\tau)^{-1}(\mathbf{A}^\tau+\mathbf{I})$. Then, Eq.~\eqref{eq:communication} can be reformulated as
\begin{equation}
    \mathbf{s}^{\tau+1} = \alpha \cdot \mathbf{d}^{\tau} + (1-\alpha) \cdot \widetilde{\mathbf{A}}^\tau \mathbf{s}^{\tau}.
    \label{eq:communication_matrix}
\end{equation}

To analyze the effect of communication, this paper makes several assumptions to simplify Eq.~\eqref{eq:communication_matrix}:
\begin{itemize}
    \item [(1)] Only the effect of communication is considered, while the dissimilarity is ignored, i.e., $\alpha=0$.
    \item [(2)] Agents are allowed to communicate multiple times within a short time period $[\tau,\tau+1]$, where $\widetilde{\mathbf{A}}^\tau$ remains unchanged.
    \item [(3)] The communication graph is connected. $\widetilde{\mathbf{A}}^\tau$ is irreducible and aperiodic.
\end{itemize}

Under the above assumptions, agents will reach consensus on detecting scores after sufficient rounds of communication,
\begin{equation}
        \mathbf{s}^{\tau+1} = \left(\underset{N\to\infty}{\text{lim}}\left(\mathbf{A}^\tau\right)^N\right)\mathbf{s}^{\tau}\\
        =\mathbf{1}\boldsymbol{\pi}^\top\mathbf{s}^{\tau}
        = C\mathbf{1},
\end{equation}
where $C=\boldsymbol{\pi}^\top\mathbf{s}^{\tau}$ is a constant. The existence of the limit is assured by the stationary distribution of the normalized random walk matrix~\cite{liu2020towards,perron1907zur}, and $\boldsymbol{\pi}\in\mathbb{R}^{|\mathcal{V}|}$ is a vector representing the stationary distribution.

However, the assumptions rarely hold in practice. The dissimilarity is non-ignorable, the communication graph is not necessarily connected, and changes frequently. Thus, it is hard to detect emergence in a fully distributed manner by utilizing the consensus of agents. Therefore, at least one global monitor is required for emergence detection.

\subsection{Brief Introduction of BYOL}
\begin{figure}[!t]
    \centering
    \includegraphics[width=0.49\textwidth]{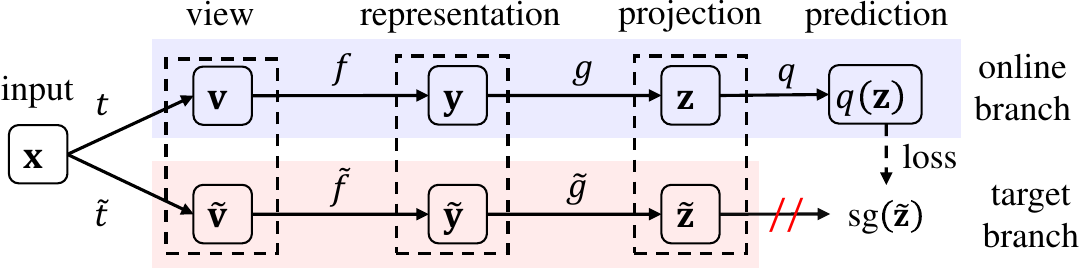}
    \caption{Architecture of BYOL~\cite{grill2020bootstrap}.}
    \label{fig:byol}
\end{figure}
BYOL~\cite{grill2020bootstrap} is a representative non-contrastive self-supervised learning method. It avoids explicit negative samples by aligning different views of the same sample encoded by asymmetric dual-branch neural networks. As shown in Figure~\ref{fig:byol}, BYOL trains two neural networks, an online network and a target network. They share the same architecture but have different parameters. Variables and functions from the target branch are denoted with a tilde, while those from the online network are without tildes.

Given a sample $\mathbf{x}$, two views are constructed via data augmentations, i.e., $\mathbf{v}=t(\mathbf{x})$ and $\widetilde{\mathbf{v}}=\widetilde{t}(\mathbf{x})$. In each branch, the view is encoded to a latent representation $\mathbf{y}$ via an encoder $f$, and then projected to be $\mathbf{z}$ via a projector $g$. The online branch learns to predict $\widetilde{\mathbf{z}}$ from the target branch via a predictor $q$. Let $\Theta$ and $\widetilde{\Theta}$ be the parameters of the online network and the target network, respectively. The loss function for the online branch is the squared error between the $\ell_2$-normalized $q(\mathbf{z})$ and $\widetilde{\mathbf{z}}$,
\begin{equation}
    \mathcal{L}\left(\Theta,\widetilde{\Theta}\right) = \left\lVert \frac{q(\mathbf{z})}{\lVert q(\mathbf{z})\rVert_2} - \frac{\widetilde{\mathbf{z}}}{\lVert \widetilde{\mathbf{z}}\rVert_2}\right\rVert^2_2.
\end{equation}
It can be shown that $\mathcal{L}(\Theta,\widetilde{\Theta})=2(1-\text{cos}(q(\mathbf{z}),\widetilde{\mathbf{z}}))$, which coincides the dissimilarity function defined in Eq.~(7) of the main text except a constant.

Exchanging the role of $\mathbf{z}$ and $\widetilde{\mathbf{z}}$ gives rise to the loss function $\widetilde{\mathcal{L}}(\Theta,\widetilde{\Theta})$ for the target branch. The final loss function is a symmetric one by summing up the losses from both branches, i.e., $\mathcal{L}=\mathcal{L}(\Theta,\widetilde{\Theta})+\widetilde{\mathcal{L}}(\Theta,\widetilde{\Theta})$.
To avoid learning collapse representations, BYOL only backpropagates through $\Theta$, and stops the gradient (sg) through $\widetilde{\Theta}$. $\Theta$ is updated by some gradient-based optimizer, while $\widetilde{\Theta}$ is update by exponential moving average, i.e.,
\begin{equation}
            \Theta \leftarrow \text{Opt}\left(\mathcal{L}\right),\quad
        \widetilde{\Theta} \leftarrow \eta \widetilde{\Theta} + (1 - \eta)\Theta,
    \label{eq:update_byol}
\end{equation}
where $\eta\in [0, 1]$ is a decay rate.

STCL differs from BYOL in several aspects:
\begin{itemize}
    \item [(1)] STCL constructs different views by leveraging the intrinsic spatio-temporal consistency of data, without hand-crafted augmentation tricks that may destroy the spatio-temporal semantics.
    \item [(2)] The asymmetric network structure is induced naturally by the spatial and temporal characteristics of data rather than being manually manipulated.
    \item [(3)] The final loss function is not symmetrized, because the symmetric variant is found to perform poorly.
\end{itemize}

\section{Summary of Notations}

Notations used in the main text are summarized in Table~\ref{tab:notations}. 
\begin{table*}[!b]
    \renewcommand{\arraystretch}{1.3}
    \caption{Summary of notations.}
    \centering
    \begin{tabularx}{\textwidth}{l|X}
        \hline
        \multicolumn{1}{l|}{Notations} & \multicolumn{1}{c}{Descriptions} \\
        \hline
        $\mathcal{G}^t=(\mathcal{V},\mathcal{E}^t,\mathbf{X}^t)$ & The interaction graph of agents at time $t$, with $\mathcal{V}$ as the set of agents, $\mathcal{E}^t$ as the set of edges, and $\mathbf{X}^t$ as the states of agents.\\
        $\mathcal{RG}^t=(\mathcal{RV},\mathcal{RE},\mathbf{y}^t)$ & The region graph at time $t$, with $\mathcal{V}$ as the set of regions, $\mathcal{E}^t$ as the set of edges, and $\mathbf{y}^t$ as the states of regions.\\
        $\mathcal{T}^*, \widehat{\mathcal{T}}$ & The set of ground truth and detected change points, respectively.\\
        \hline
        $\mathcal{N}^t_j=\{i:d^t_{ij}\leq \delta\}$ & The set of agent $j$'s neighbors at time $t$ with a radius of $\delta$.\\
        $[\tau(w),\tau]$ & The time window for recording agents' states. $w$ is the window size and $\tau(w)=\tau-w+1$. \\
        $\mathbf{x}^t_j$ & State of agent $j$ at time $t$.\\
        $\mathbf{z}^t_j,\mathbf{h}^t_j$ & Agent representations from the spatial transformer and the temporal transformer, respectively.\\
        $\mathbf{v}^t_j,\mathbf{v}^{(\tau)}_j$ & The transient and short-term representation for evaluating agent-level temporal consistency loss.\\
        $\mathbf{n}^t_j,\mathbf{m}^{(\tau)}_j$ & The transient and short-term representation for evaluating agent-level spatial consistency loss.\\
        $\text{Proj}_S,\text{Proj}_T$ & Agent-specific spatial and temporal projection implemented by MLP.\\
        $\text{Pool}_S,\text{Pool}_T$ & Agent-specific spatial and temporal pooling, respectively.\\
        $\text{STE}_A$ & Agent-level spatio-temporal encoder.\\
        \hline
        $\mathbf{r}^t_m$ & Region $m$'s representation at time $t$.\\
        $\mathbf{r}^t_\mathcal{G},\mathbf{u}^{(\tau)}$ & The transient and short-term representation for evaluating system-level temporal consistency loss.\\
        $\mathbf{w}^{(\tau)}_m$ & The short-term region representation for evaluating system-level spatial consistency loss.\\
        $\text{Proj}_{RS},\text{Proj}_{RT}$ & Regional spatial and temporal projection implemented by MLP.\\
        $\text{Pool}_{RS},\text{Pool}_{RT}$ & Regional spatial and temporal pooling, respectively.\\
        $\text{STE}_R$ & Region-level spatio-temporal encoder.\\
        \hline
        $\alpha$ & Mixing coefficient for the communication of agents.\\
        $s^t_j,s^t_\mathcal{G}$ & Agent-level and system-level detecting score, respectively.\\
        \hline
        $\kappa$ & Number of positive pairs for computing the spatial consistency loss.\\
        $\mathcal{L}_{T}, \mathcal{L}_{S}, \mathcal{L}_\text{Agent}$ & Agent-level temporal consistency loss, spatial consistency loss and overall loss, respectively.\\
        $\mathcal{L}_{ST}, \mathcal{L}_{SS}, \mathcal{L}_\text{System}$ & System-level temporal consistency loss, spatial consistency loss, and overall loss, respectively.\\
        \hline
    \end{tabularx}
    \label{tab:notations}
\end{table*}

\section{Algorithmic Details of HSTCL}
\begin{algorithm*}[!b]
    \caption{Agent-Level STCL}
    \LinesNumbered
    \label{alg:agent_stcl}
    \KwIn{States of agents $\mathbf{X}^{1:T}$ from $S$ times of simulation, size of time window $w$, number of positive pairs $\kappa$ for spatial consistency loss, number of training epochs $E$, number of selected samples $B$ for each simulation}
    \KwOut{Parameters of the online and target networks, $\Theta_A$ and $\widetilde{\Theta}_A$}

    Initialize $\Theta_A$ and $\widetilde{\Theta}_A$ with random weights;

    \For{$e=1:E$}{
        \ForEach{simulation}{
            \For{$l=1:B$}{
                Select a random slice of agent states $\mathbf{X}^{\tau(w):\tau}$;

                Compute the agent representations via Eqs.~(2)-(5);

                Compute the temporal consistency loss via Eqs.~(6)-(8);
                
                Compute the spatial consistency loss via Eqs.~(9)-(10);

                Compute the total loss via Eq.~(11) and update $\Theta_A$ and $\widetilde{\Theta}_A$ via Eq.~(12);
            }
        }
    }
\end{algorithm*}
\begin{algorithm*}[!t]
    \caption{System-Level STCL}
    \LinesNumbered
    \label{alg:system_stcl}
    \KwIn{States of regions $\mathbf{y}^{1:T}$ from $S$ times of simulation, size of time window $w$, number of positive pairs $\kappa$ for spatial consistency loss, number of training epochs $E$, number of selected samples $B$ for each simulation}
    \KwOut{Parameters of the online and target networks, $\Theta_R$ and $\widetilde{\Theta}_R$}

    Initialize $\Theta_R$ and $\widetilde{\Theta}_R$ with random weights;

    \ForEach{simulation}{
        Aggregate agent-level detecting results into region states via Eq.~(14);
    }

    \For{$e=1:E$}{
        \ForEach{simulation}{
            \For{$l=1:B$}{
                Select a random slice of region states $\mathbf{y}^{\tau(w):\tau}$;

                Compute the region representations via Eq.~(15);
                
                Compute the temporal consistency loss via Eqs.~(16)-(18);
                
                Compute the spatial consistency loss via Eqs.~(19)-(20);

                Compute the total loss via Eq.~(21) and update $\Theta_R$ and $\widetilde{\Theta}_R$ via Eq.~(12);
            }
        }
    }
\end{algorithm*}
\begin{algorithm*}[!t]
    \caption{Procedure of Emergence Detection}
    \LinesNumbered
    \label{alg:emergence}
    \KwIn{States of agents $\mathbf{X}^{1:T}$, size of time window $w$, mixing coefficient $\alpha$ for agents' communication, threshold $c$ for change-point detection
    }
    
    \KwOut{A set of detected time steps of emergence formation and evaporation $\widehat{\mathcal{T}}$}

    \For{$\tau=1:T$}{
        \eIf{$\tau<w$}{
            \tcp{Initialization}
            \ForEach{agent $j$}{
                Set agent-level detection score $s^\tau_j=0$;

                Record the states of neighbors $\{\mathbf{x}^\tau_i:i\in\mathcal{N}^\tau_j\}$;
            }
            Set the state $y^\tau_m=0$ for each region $\mathcal{R}_m$;

            Set system-level detection score $s^\tau_\mathcal{G}=0$;
        }{
            \tcp{Agent-level detection}
            \ForEach{agent $j$}{
                Read the states of itself and neighbors during the last time window, i.e., $\mathbf{x}^{\tau(w):\tau}$ and $\{\mathbf{x}^t_i:i\in\mathcal{N}^t_j\}^{\tau}_{t=\tau(w)}$;

                Obtain agent representations $\mathbf{h}^{\tau(w):\tau}_j$ via Eqs.~(2)-(5);

                Compute the detecting score $s^\tau_j$ via Eq.~(13);
            }

            \tcp{System-level detection}
            \ForEach{region $\mathcal{R}_m$}{
                Collect agent-level detection scores and aggregate them to the region state $y^\tau_n$ via Eq.~(14);

                Obtain region representations $\mathbf{r}^{\tau(w):\tau}_m$ via Eq~(15);

                Obtain the system representation $\mathbf{u}^{(\tau)}$ via Eqs.~(16)-(17);
            }
            \tcp{Detect time steps of emergence formation and evaporation using the criterion defined in Definition~4}

            Compute the system-level detection score $s^\tau_\mathcal{G}$ via Eq.~(22);
            
            \If{$s^{\tau-1}_\mathcal{G}>c$ and $s^\tau_\mathcal{G}\leq c$}{
                Add $\tau-1$ to the set $\widehat{\mathcal{T}}$;
            }
        }
    }
\end{algorithm*}

\subsection{Incremental Computation of The STE}
This subsection analyses how the STE can update the spatial and the temporal representations incrementally. From Eq.~(3) of the main text, the spatial representations of agents are computed independently at each time step. Thus, for time window $[\tau(w)+1, \tau+1]$, only $\mathbf{z}^{\tau+1}$ is needs to be computed, while $\mathbf{z}^{\tau(w)+1:\tau}$ from the time window $[\tau(w),\tau]$ can be reused. The extra computational cost is $O(|\mathcal{N}^{\tau+1}|)$, which is irrelevant to the window size $w$.

Similarly, for the temporal representations, $\mathbf{q}^{\tau(w)+1:\tau}$, $\mathbf{k}^{\tau(w)+1:\tau}$ and $\mathbf{v}^{\tau(w)+1:\tau}$ can be reused. Let
\begin{equation}
    \mathbf{W}^\tau = \mathbf{q}^{\tau(w):\tau}\left(\mathbf{k}^{\tau(w):\tau}\right)^\top
    \label{eq:temporal_attention}
\end{equation}
be the unnormalized attention matrix. Then,
\begin{equation}
    \mathbf{W}^{\tau+1}=
    \begin{bmatrix}
        \mathbf{W}^{\tau}[\tau(w)+1:\tau,\tau(w)+1:\tau] & \mathbf{q}^{\tau(w)+1:\tau} \mathbf{k}^{\tau+1}\\
        \left(\mathbf{q}^{\tau(w)+1:\tau}\mathbf{k}^{\tau+1}\right)^\top & \left(\mathbf{q}^{\tau+1}\right)^\top \mathbf{k}^{\tau+1}
    \end{bmatrix}.
\end{equation}
The extra computational cost is $O(w)$. Thus, the incremental update is more efficient than the naive calculation of $\mathbf{W}^{\tau+1}$ with a time complexity of $O(w^2)$.

Given the unnormalized attention matrix, the temporal representations can be updated as
\begin{equation}
    \mathbf{h}^{\tau(w)+1:\tau+1}=\text{softmax}\left(D^{-\frac{1}{2}}\mathbf{W}^{\tau+1}\right)
    \begin{bmatrix}
        \mathbf{v}^{\tau(w)+1:\tau}\\
        \left(\mathbf{v}^{\tau+1}\right)^\top
    \end{bmatrix}.
\end{equation}

\subsection{Pseudo Code}
The pseudo codes of agent-level STCL, system-level STCL, and emergence detection are shown in Algorithm~\ref{alg:agent_stcl}, Algorithm~\ref{alg:system_stcl} and Algorithm~\ref{alg:emergence}, respectively.

\section{Experimental Details}
\begin{figure*}[!t]
    \begin{minipage}{0.32\textwidth}
        \centering
        \includegraphics[width=\textwidth]{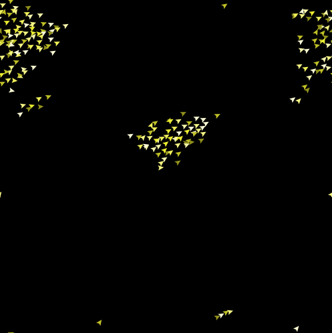}
        \caption{Flocking on the Flock dataset. Birds are in yellow and patches containing no birds are in black.}
        \label{fig:flocking}
    \end{minipage}
    \hfil
    \begin{minipage}{0.32\textwidth}
        \centering
        \includegraphics[width=\textwidth]{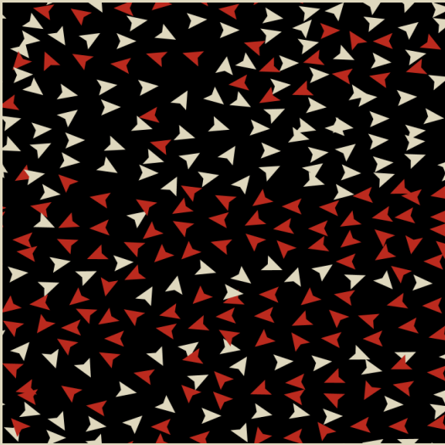}
        \caption{Counter-flow on the Pedestrian dataset. Pedestrians in white walk from left to right, and those in red walk from right to left. Pedestrians walking in the same directions form a lane.}
        \label{fig:counter-flow}
    \end{minipage}
    \begin{minipage}{0.32\textwidth}
        \centering
        \includegraphics[width=\textwidth]{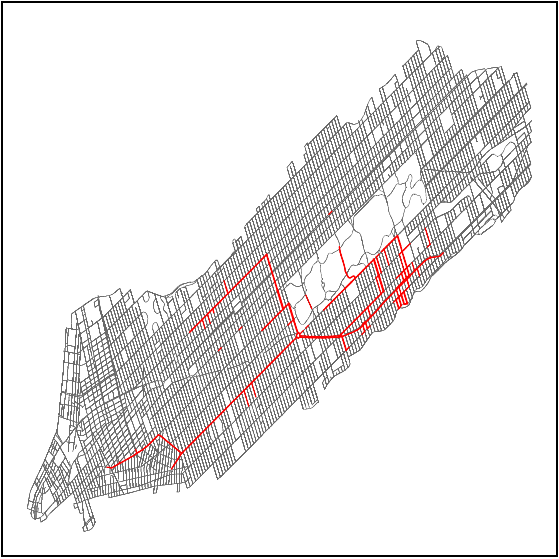}
    \caption{Network-level congestion on the Traffic dataset. Congested road segments are in red, and the normal ones are in gray.}
    \label{fig:congestion}
    \end{minipage}
\end{figure*}
\begin{figure*}[!t]
    \begin{minipage}{0.32\textwidth}
        \centering
        \includegraphics[width=\textwidth]{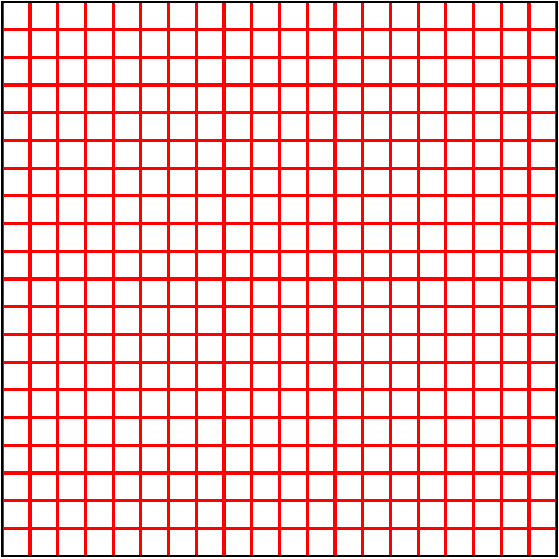}
        \caption{$20\times 20$ regions covering the area of interest on the Flock and Pedestrian datasets.}
        \label{fig:flock_grid}
    \end{minipage}
    \hfil
    \begin{minipage}{0.32\textwidth}
        \centering
        \includegraphics[width=\textwidth]{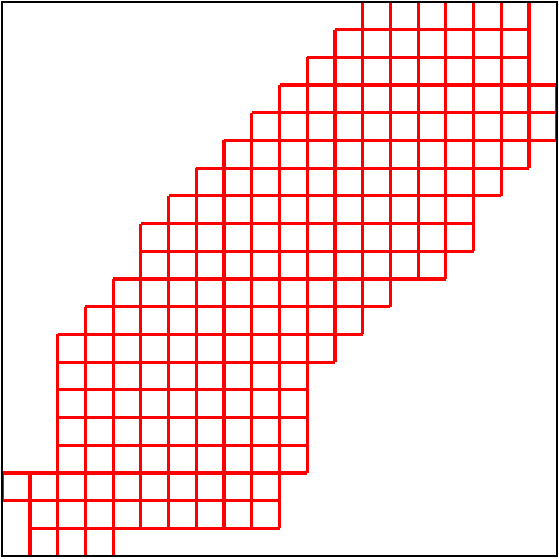}
        \caption{Regions covering the road net on the Traffic dataset.}
        \label{fig:traffic_grid}
    \end{minipage}
    \hfil
    \begin{minipage}{0.32\textwidth}
        \centering
        \includegraphics[width=0.6\textwidth]{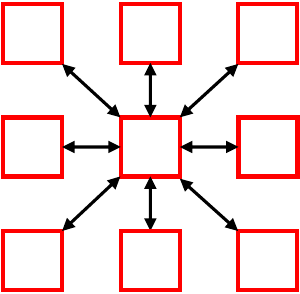}
        \caption{Construction of the region graph. Each region is a node, and bidirectional edges exist between two adjacent regions.}
        \label{fig:grid_graph}
    \end{minipage}
\end{figure*}
\subsection{Visualizations of Emergent Behaviors on All Datasets}
To gain an intuitive understanding of the emergent behaviors on the Flock, Pedestrian, and Traffic datasets, this paper visualizes them in Figure~\ref{fig:flocking}-\ref{fig:congestion}.

\subsection{Additional Details of Traffic Dataset}
The Traffic dataset tries to recover the true traffic flow by combining the observational data and simulation rules. The road net is taken from the real world. The simulation of DETect relies on a file that contains 131,559 records of 6,500 unique cars. Each record contains the car ID, the longitudes and latitudes for the starting spot and the ending spot, the distance between two spots, and the duration. Contiguous records of the same car form the path that it travels. The waypoints between two adjacent spots are unobserved and are approximated by the routing algorithm provided by Graphhopper~\cite{graphhopper}. The movement of a car is determined by the simulation rules constrained by the route.

\subsection{Implementation Details of HSTCL}
The dimensions of all hidden layers of models are 128. All experiments are run 5 times on a machine containing 128GB of RAM, and 8 NVIDIA RTX2080Ti graphics cards with PyTorch 1.9.1 and CUDA 11.1 in Ubuntu 20.04.

To stay consistent with DETect, this paper sets key hyperparameters as follows. The state series of agents are downsampled every 5 steps. The radius $\delta$ for defining an agent's neighborhood is set to 5 on the Flock and Pedestrian datasets, and 10 on the Traffic dataset. The mixing coefficient $\alpha$ for communication is set to 0.05.

For system-level detection, the area where all agents move is split into a $20\times 20$ grid, resulting in 400 regions. On Traffic dataset, regions that cover no streets are removed, resulting in 186 regions. In the region graph, bidirectional edges exist between two adjacent regions. The partitions of regions on the Flock and Pedestrian datasets are shown in Figure~\ref{fig:flock_grid}, and the partition of regions on the Traffic dataset is shown in Figure~\ref{fig:traffic_grid}. The construction of the region graph is shown in Figure~\ref{fig:grid_graph}. More sophisticated strategies can be designed for region partition, e.g., regions with irregular shapes and sizes, which is left for future work.

The states of all agents or regions within a time window are regarded as a single sample. All methods are trained for 10 epochs. The thresholds for both agent-level and system-level detection scores are determined by grid search on the validation set w.r.t. F1 score. The size of time window is set to $10$ for agent-level detection, and $40$ for system-level detection. A discussion is shown in Section~IV-F of the main text.

\subsection{Visualization of Edge Counts Over Time}
\begin{figure}[!t]
    \centering
    \includegraphics[width=0.99\columnwidth]{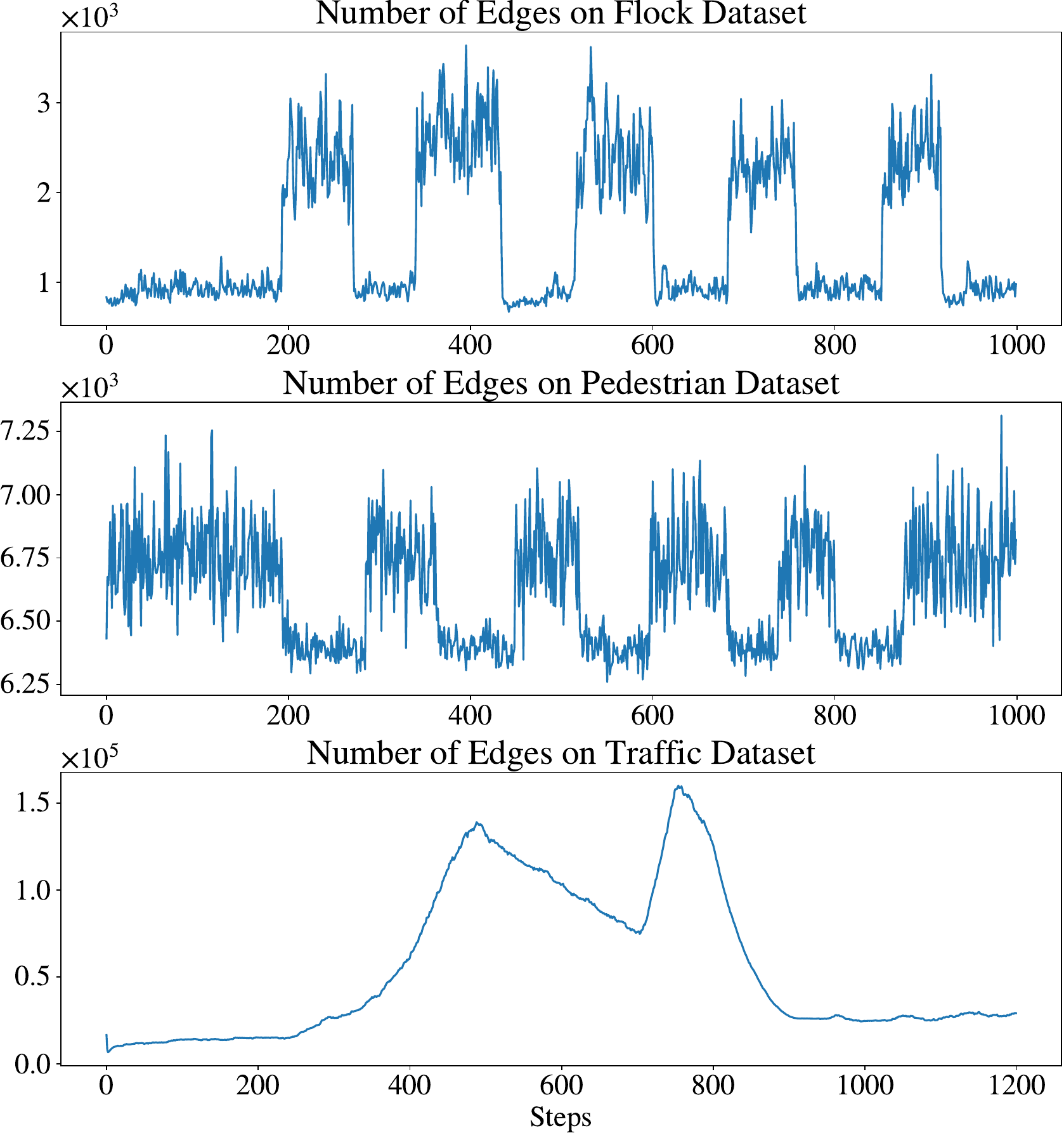}
    \caption{Numbers of edges over time on different datasets.}
    \label{fig:edge_count}
\end{figure}

Since agents' positions change over time, the edges of the interaction graphs are essentially dynamic. To verify this fact, this paper randomly selects a simulation for each dataset and visualizes the number of edges over time in Figure~\ref{fig:edge_count}. The results confirm the frequent addition and removal of edges over time. It is worthwhile to point out that the edge count is not always a reliable statistic for emergence detection. On Flock and Pedestrian datasets, the edge counts even fluctuate during non-emergent periods. When only limited information within a short time window is available, the edge count is insufficient for online detection. On Traffic dataset, the edge count is less sensitive to emergence formation and evaporation. Indeed, DETect has selected the node degree, a closely related statistic, as the feature, but still fails to detect emergence accurately.

\subsection{Difference Between F1 Score and Covering metric}\label{sec:exp:f1_vs_cover}
\begin{figure}[!t]
    \centering
    \includegraphics[width=0.99\columnwidth]{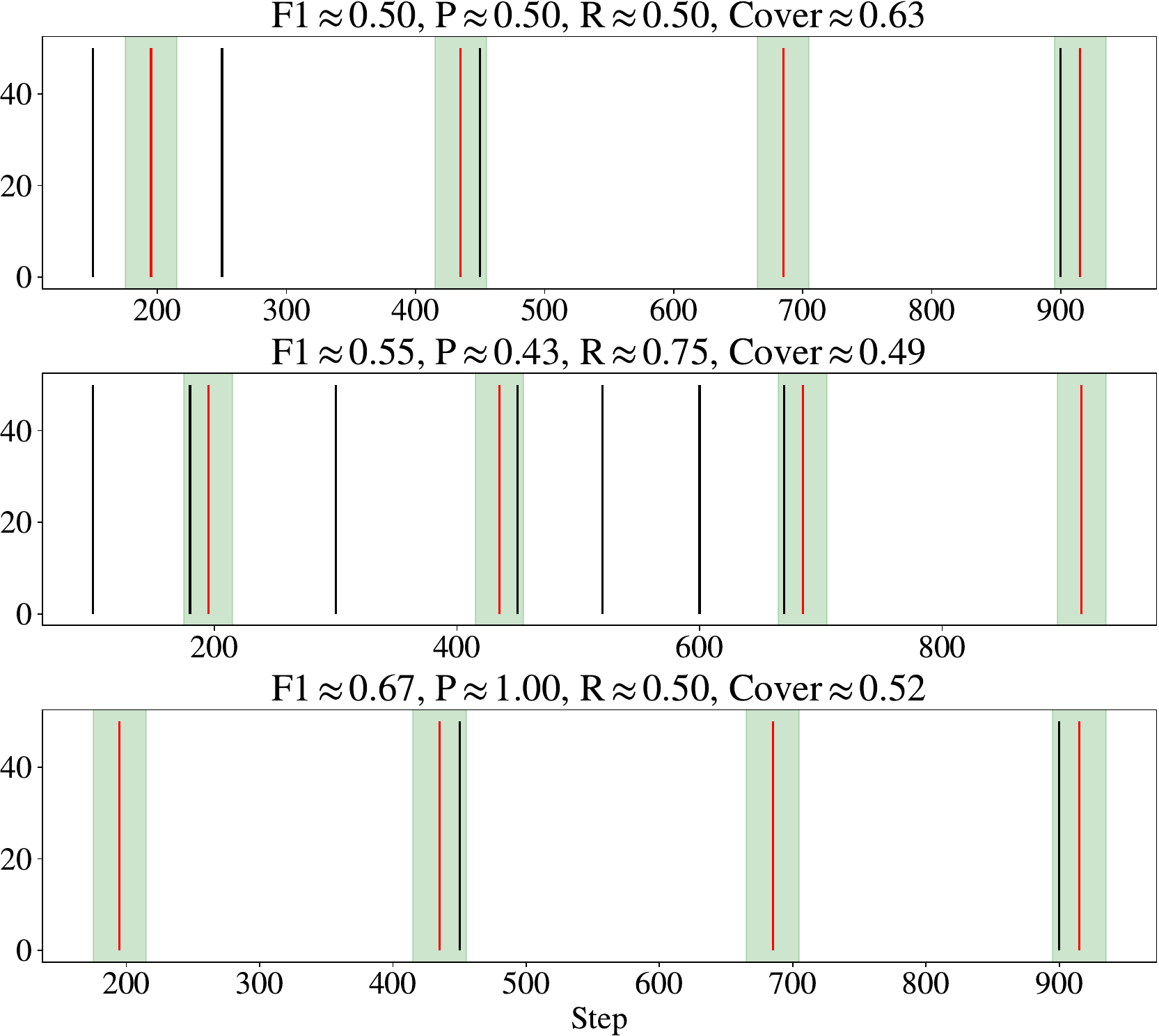}
    \caption{Examples of detected results with higher F1 scores and lower covering metrics. The \textit{first} row is regarded as a reference, the \textit{second} row is an example with a higher recall rate and a lower covering metric, and the \textit{third} row is an example with a higher precision and a lower covering metric. The ground truth change points and the detected ones are in red lines and black lines, respectively. The span within the margin of a change point is in green. Best viewed in color.}
    \label{fig:f1_vs_covering_metric}
\end{figure}

The F1 score and covering metric are two different metrics. The F1 score is the harmonic mean of precision (P) and recall rate (R). It is sensitive to the threshold $\theta$ that measures the error tolerance between the detected and the ground truth change points, and is generally irrelevant to the length of each segment. By contrast, the covering metric is irrelevant to $\theta$ but is sensitive to the partition of segments.
Hence, they are not always positively correlated. A higher F1 score can be accompanied by a lower covering metric. This paper handcrafts three detecting results and visualizes them in Figure~\ref{fig:f1_vs_covering_metric} to demonstrate the possibility. The second row increases the F1 score by sacrificing the precision, resulting in multiple short segments. According to the definition of the covering metric, for a given ground truth segment, only the most overlapping detected segments will contribute to the covering metric. Thus, fragmented detecting results will lead to a lower covering metric. The third row increases the F1 score by sacrificing the recall rate, which means some detected segments can be much longer than their corresponding ground truth segment, resulting in a lower covering metric.

\bibliographystyle{IEEEtranN}
\bibliography{IEEEabrv,ref}